\documentclass[a4paper,fleqn,usenatbib]{mnras}
\usepackage[T1]{fontenc}
\usepackage{ae,aecompl}

\usepackage{subfig}
\usepackage{graphicx}	% Including figure files
\usepackage{amsmath}	% Advanced maths commands
\usepackage{amssymb}	% Extra maths symbols

\title[A near-core convective shell in sdB models]{The effects of near-core
  convective shells 
on the gravity modes of the subdwarf B pulsator KIC\,10553698A
}

\author[H.Ghasemi et al.]{
H. Ghasemi,$^{1,2}$\thanks{E-mail: h.ghasemi@znu.ac.ir}
E. Moravveji,$^{2}$\thanks{Marie Curie postdoctoral fellow}
C. Aerts,$^{2,3}$
H. Safari,$^{1}$
M. Vu\v{c}kovi\'{c}$^{4}$
\\
$^{1}$Department of Physics, University of Zanjan, P.O. Box 45195-313, Zanjan, Iran\\
$^{2}$Institute of Astronomy,  Celestijnenlaan 200D, 3001 Leuven, Belgium\\
$^{3}$Department of Astrophysics, IMAPP, Radboud University Nijmegen, 6500 GL, Nijmegen, The Netherlands\\
$^{4}$Instituto de F\'{i}sica y Astronom\'{i}a, Facultad de Ciencias, Universidad de Valpara\'{i}so, Gran Breta\~{n}a 1111, Playa Ancha, Valpara\'{i}so 2360102, Chile \\}
\date{Accepted 2016 October 31. Received 2016 October 27; in original form 2016 October 8}

\pubyear{2016}

\begin{document}
\label{firstpage}
\pagerange{\pageref{firstpage}--\pageref{lastpage}}
\maketitle

% Abstract of the paper
\begin{abstract}
KIC 10553698A is a hot pulsating subdwarf B (sdB) star observed by the Kepler 
satellite. It exhibits dipole $(l=1)$ and quadrupole $(l=2)$ gravity modes 
with a clear period spacing structure. The seismic properties of the KIC 10553698A
provide a test of stellar evolution models, and offer a unique opportunity to determine 
mixing processes. We consider mixing due to convective overshooting beyond
the boundary of the helium burning core. Very small overshooting ($f = 10^{-6}$) results
in a progressive increase in the size of convective core. However, moderate ($f = 10^{-2}$)
and small ($f = 10^{-5}$) overshooting both lead to the occurrence of inert outer convective
shells in the near-core region. We illustrate that the chemical stratifications induced 
by convective shells are able to change the g-mode period spacing
pattern of a sdB star appreciably. The mean period spacing and trapping of the gravity modes 
in the model with moderate and small core overshooting are fully consistent
with the period-spacing trends observed in KIC 10553698A.  Atomic diffusion driven
by gravitational settling as well as thermal and chemical gradients is applied to reach
a better match with the observed period spacings. Models that include small or very
small overshooting with atomic diffusion have a decreased lifetime of the extreme 
horizontal branch phase and produce chemical stratification induced by convective shells
during helium burning phase. In addition of being consistent with asteroseismology,
their calculated values of the $R_2$ parameter are more compatible with the observed $R_2$ values.

\end{abstract}

% Select between one and six entries from the list of approved keywords.
% Don't make up new ones.

%%%%%%%%%%%%%%%%%%%%%%%%%%%%%%%%%%%%%%%%%%%%%%%%%%

%%%%%%%%%%%%%%%%% BODY OF PAPER %%%%%%%%%%%%%%%%%%

\section{Introduction}\label{s-intro}

Subdwarf B (sdB) stars are low mass core-helium burning objects situated at the extreme
horizontal branch (EHB).  Close binary interaction and Roche Lobe Overflow (RLOF) on
the red giant branch are two essential ingredients of the formation scenarios
that strip off the outer envelope and produce a sdB star
\citep[e.g.,][]{2009ARA&A..47..211H}.  Short period binary systems form as a
consequence of common envelope interaction while stable RLOF produces wider binary systems.  
The sdB stars lose almost their entire outer hydrogen-rich envelope and occur on the EHB.

In core He burning models, the mixing between CO-rich and He-rich layers changes
the chemical discontinuity between the fully mixed core and the stable radiative
envelope. This increases the opacity beyond the convective boundary. Therefore,
during the horizontal branch phase, the helium core size can increase or stay
unchanged depending on the details of the adopted mixing. In fact, the detailed
prescription for near-core mixing phenomena has been debated in the literature
over the past decades.  As an example, semiconvection was proposed as a
plausible mixing mechanism in the radiative interior of core He-burning stars by
\citet{1985ApJ...296..204C}, while \citet{1990ASPC...11....1S} excluded it.  
Alternatively, mixing due to core
overshooting by the penetration of convective eddies from the fully convective
core into the radiatively stable regions was also proposed for core-He burning
models \citep[see][and references
therein]{2015MNRAS.452..123C,2015MNRAS.453.2290B}. As a result, the properties
of He burning cores remain largely unknown.

The shape and efficiency of near-core mixing is connected with the 
ratio $R_2$, which stands for the number of stars observed on the early
asymptotic giant branch versus the number of stars in the horizontal
branch (HB) phase \citep[e.g.,][for a review]{1988ARA&A..26..199R}.  The traditional
canonical evolutionary theory for the HB, relying on semiconvection and
suppressing the so-called breathing pulses near the end of the HB phase, was
found to be compatible with the measured values of $R_2= 0.15 \pm 0.01$ in
globular clusters \citep[e.g.,][]{1985A&A...145...97B,2001A&A...366..578C}.
\cite {2016MNRAS.456.3866C} recently obtained $R_2= 0.117 \pm 0.005$ from observations based on 
different criteria to restrict the counts of AGB stars.

We argue that asteroseismology of pulsating compact stars offers
a new window on the discrimination between different mixing mechanisms, thanks to
the high sensitivity of the pulsation frequencies to the details of the input
physics of evolutionary models \citep[e.g.,][]{1994ApJ...427..415K}.  
So far, three classes of pulsating sdB stars were recognized: 
\begin{enumerate}
\item EC\,14026 pulsators exhibit short-period pulsations first discovered by
  \citet{1997MNRAS.285..640K} and corresponding to theoretical low-degree
  pressure (p-) modes driven by the ${\kappa}$\,mechanism 
  due to the iron opacity bump boosted by radiative
  levitation \citep{1996ApJ...471L.103C}. The dominant restoring force for
  these rapid p-modes is connected with the pressure gradient.

\item PG\,1716 pulsators exhibit long-period pulsations discovered by
  \citet{2003ApJ...583L..31G}.  These slow pulsations are interpreted as gravity
  (g-) modes \citep{2003ApJ...597..518F}. The dominant restoring force for these
  long-period instabilities is the buoyancy force, which depends on the detailed
  internal structure of the sdB. The theoretical instability strip for g-modes 
  excited by the ${\kappa}$ mechanism operating in the iron opacity bump around 
  $\log T\sim5.2$ is consistent with observations \citep[][and references therein]
  {2014A&A...569A.123B}.
  
\item Hybrid sdB stars demonstrate both p-modes and g-modes simultaneously \citep{2006A&A...445L..31S}.
  
\end{enumerate}

The detection and identification of long-period low-amplitude g-modes from the ground 
based data is challenging. However, the high-precision space observations of sdB pulsators 
with the CoRoT and Kepler satellites has facilitated in-depth seismic modelling of these 
class of pulsating stars. A breakthrough in linking the pulsations with the size of 
convective cores in sdB stars has occurred by the studies of \citep{2010A&A...524A..63V,2010ApJ...718L..97V,2010HiA....15..357C,2011A&A...530A...3C}.

In sdB stars, g-modes propagate from the surface all the way down to the boundary of
the convective core, because the envelopes of these stars are predominantly
radiative. Therefore, asteroseismology of sdB stars based on high-precision
space photometry is one of the best tools to test various mixing scenarios on
top of convective core. This is the method we adopt in this work.

In this paper, we use measured period spacings of the selected g-mode sdB
pulsator KIC\,10553698A as a laboratory to test the efficiency of mixing due to
convective overshooting at the edge of the convective core. Our evolutionary
and asteroseismic models are computed with the publicly available code MESA
\citep[Modules for Experiment in Stellar
Astrophysics][version\,7385]{2011ApJS..192....3P,2013ApJS..208....4P,2015ApJS..220...15P},
in combination with the GYRE linear adiabatic pulsation code
\citep[][version\,4.0]{2013MNRAS.435.3406T}.

We compare the evolutionary and asteroseismic properties of models with
inefficient and moderately efficient overshooting with the observations. We
find that models with moderate and small core overshooting lead to inert convective shells
adjacent to the convective core. These shells can effectively contribute to the
trapping of g-modes. The period spacing pattern of KIC\,10553698A provides
evidence in support of the presence of such a convective shells.

Equipped with the measured period spacing pattern of its g-modes, we investigate
the effect of atomic diffusion as an extra mixing process beyond the convective
core, which leads to different behaviour of the convective
core\cite[e.g.,][]{2015ApJ...806..178S} and better match with observed period
spacing.  Different kinds of mixing due to core overshooting or atomic diffusion
in horizontal branch (HB) stars lead to various shapes of the growing convective
core, with similar EHB evolution scenarios.  We consider the $R_2$ parameter as
a second test of the nature of the overshooting beyond the growing inner core
for the core helium burning stars.

\section{The case of the sdB pulsator KIC\,10553698A}\label{s-obs}

The evenly-spaced long-cadence high-quality {\it {\it Kepler\/}} data provide an
excellent opportunity to study low-frequency g-mode pulsations, for which
ground-based surveys have hardly provided seismic solutions. As examples, the
uninterrupted nature of these data opened the discovery of g-mode period
spacings of pulsators in the core-hydrogen burning phase whose seismic modelling
was impossible from the ground
\citep{2014A&A...570A...8P,2015ApJ...803L..25P,2015ApJS..218...27V,2015A&A...584A..35S}
and illustrated the requirement of chemical mixing in the near-core regions of
such stars
\citep{2015A&A...580A..27M,2016ApJ...823..130M,2016arXiv160700820V,2016A&A...592A.116S}.
The high-order g-mode frequencies in sdB stars required short cadence {\it
  Kepler\/} data with 58.5\,s integration times. Observational investigations of
{\it {\it Kepler\/}} sdB stars have been published in a series of papers, e.g.,
\citet{2010MNRAS.409.1470O, 2010MNRAS.409.1487K, 2010MNRAS.409.1496R,
  2010MNRAS.409.1509K, 2011MNRAS.414.2860O, 2011MNRAS.414.2871B,
  2011MNRAS.414.2885R, 2012ApJ...753L..17O, 2012A&A...544A...1T,
  2012MNRAS.424.2686B, 2012MNRAS.427.1245R, 2014A&A...564L..14O,
  2014MNRAS.440.3809R, 2014A&A...570A.129T, 2015ApJ...805...94F,
  2016A&A...585A..66B}.  Here, we focus on the target with the highest potential
for near-core structure probing by trapped g-modes.

KIC\,10553698A was observed with the short-cadence mode of {\it Kepler\/} for
most of the duration of the mission and turned out to be a rich g-mode pulsator
orbiting the white dwarf KIC\,10553698B in a 3.38\,d orbit. This binary was
extensively studied by \citet{2014A&A...569A..15O}, who analysed the light curve
and detected 156 significant frequencies identified as dipole $(l=1)$ and
quadrupole $(l=2)$ modes.  The observed period spacing for dipole and quadrupole
modes provide a clear indication of mode trapping, revealed by its unambiguous
large deviations from the uniform spacing \citep[Fig.\,10
in][]{2014A&A...569A..15O}.  This structure of g-mode period spacings is
sensitive to the density and chemical stratifications between the convective
core and the radiative envelope \citep{2000ApJS..131..223C,2002ApJS..139..487C,2002ApJS..140..469C}.

\section{Computation of Stellar Models}\label{s-method}
We used the MESA code to evolve a model with 1.5\,M$_\odot$ initial mass from the pre-main sequence
phase until the core He depletion, with the following input physics.
The chemical composition was taken from the \citet{2009ARA&A..47..481A} solar
mixture (abbreviated as A09).  The initial mass fractions of H, He and heavy
elements were set to $X_{\rm ini}$ = 0.738, $Y_{\rm ini}$= 0.248, and $Z_{\rm
  ini}$= 0.014, respectively.  The Schwarzschild criterion for convective
instability was used, with the mixing length parameter fixed to the value
$\alpha_{\rm MLT}$= 1.8.  The default equation of state for Helmholtz free
energy was used, with the density and temperature as independent variables.  Due
to the critical role of local carbon and oxygen abundance on the behavior of the
core boundary, we employ the OPAL Type\,II opacity tables
\cite{1996ApJ...464..943I}, which consider the fractional abundance of carbon
and oxygen.  The adopted nuclear burning networks cover the hot CNO reactions,
the triple alpha reaction and carbon/oxygen burning, in addition to successive
alpha captures and the weak nuclear interactions.

MESA calculates the neutrino energy loss rates produced by nuclear neutrinos due
to weak interactions and thermal neutrinos generated by a range of processes
including plasmon decay, pair annihilation, Bremsstrahlung, recombination and
photo-neutrinos (Compton scattering).  This consistent treatment of the input
physics allows MESA to model off-center helium flashes in the
electron-degenerate cores of low mass stars
\citep{2011ApJS..192....3P,2012ApJ...744L...6B}.  Rotation was neglected during
the entire evolution.  Mass loss was ignored during the core hydrogen and core
helium burning phases while a Reimers' wind with $\eta$ = 0.1 \citep[][for a
definition]{2011ApJS..192....3P,2012ApJ...744L...6B} was used during the red
giant branch phase.

Mixing due to an exponential diffusive overshoot description with coefficient
$D_{\rm ov}$ was applied to all convective boundaries during all evolutionary
stages using the prescription given by \cite{1996A&A...313..497F} and
\cite{2000A&A...360..952H}:
\begin{equation}
D_{\rm ov} = D_{\rm conv}  \exp\left(\frac{-2\,z}{f\, H_{p}}\right),
\end{equation}
where $f$ is the overshooting parameter and $H_{p}$ denotes the pressure scale
height at the convective boundary.  The outward distance from the edge of the
convective zone is denoted by $z$.  $D_{\rm conv}$ is the convective mixing
diffusion coefficient, taken as $f_0 H_p$ inside the convective boundary, with
$f_0=f/10$.  No special algorithms were used to handle semiconvection or
suppress breathing pulses. The reason is that the effects of semiconvection are washed out when overshooting is
considered in evolutionary computations.  Atomic diffusion driven by
gravitational settling, as well as by temperature and chemical gradients was
included in one of the evolutionary scenarios, adopting the diffusion
coefficients from \cite{1994ApJ...421..828T}.

\section{Time steps}

MESA offers a time step selection algorithm based on absolute or relative
changes in the physical quantities throughout the evolution of a star.  MESA
uses mixing-length theory to treat the convective mixing
\citep{2011ApJS..192....3P}.  The mixing length theory determines the convective
boundary before solving the coupled structure, nuclear burning and chemical
mixing equations. An exponential diffusive core overshoot pattern leads to a
high mixing coefficient in the first layers beyond the convective core that is
almost equal to the one in the convective core itself. Consequently, these
layers become convective even with a small time step during the helium burning
phase. In the next time step, the equations should solve for a larger convective
core and a different temperature distribution.  After several consecutive small
time steps, the convective core evolution is distinct from the case where one
adopts a single large time step equal to the total of the smaller time steps.
The behaviour of the convective core is clearly sensitive to the number of time
steps taken during the horizontal branch. In general terms, smaller time steps
are required when adopting large core overshooting values when studying the
evolution of the convective core.  In this paper, all evolutionary scenarios are
done by adopting {\tt max\_years\_for\_timestep = 5000 yr} as an upper limit for
the time steps while using {\tt varcontrol\_target = 10$^{-6}$} to control
relative variations in the interior structure from one stellar model to the
next.

\section{Evolutionary aspects}\label{s-evol}

\begin{figure}
\includegraphics[width=\columnwidth]{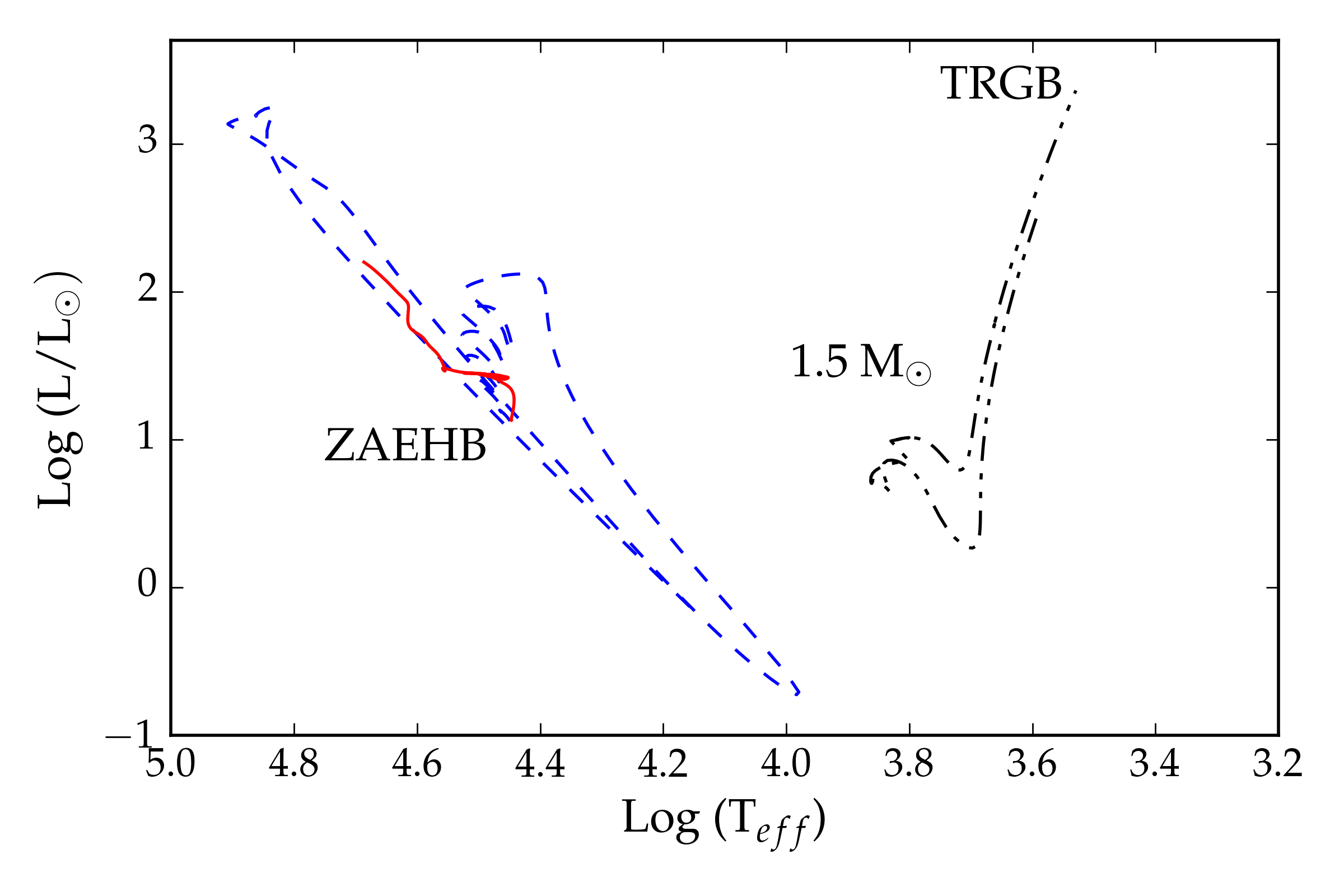}
\caption{The evolutionary track of a 1.5\,M$_\odot$ star.  The evolution from
  the pre-main-sequence to the tip of the RGB is plotted by a black dash-dot
  line.  The helium flash phase is plotted by the blue dashed line, while the
  red solid line illustrates the extreme horizontal branch phase.}
\label{fig:hr}
\end{figure}

MESA is able to evolve a model of a single (sdB) star from the pre-main sequence
phase to the tip of the red giant branch (TRGB).  While a star ascends the red
giant branch, the core contracts and the mean separation between the particles
decreases.  Eventually, the mean separation between electrons becomes of order
the de Broglie wavelength, and the helium core becomes electron-degenerate,
while the gravitational energy released from the core expands the envelope.
Fig.\,\ref{fig:hr} illustrates the evolutionary track of our 1.5\,M$_\odot$
model.

\begin{figure}
\includegraphics[width=\columnwidth]{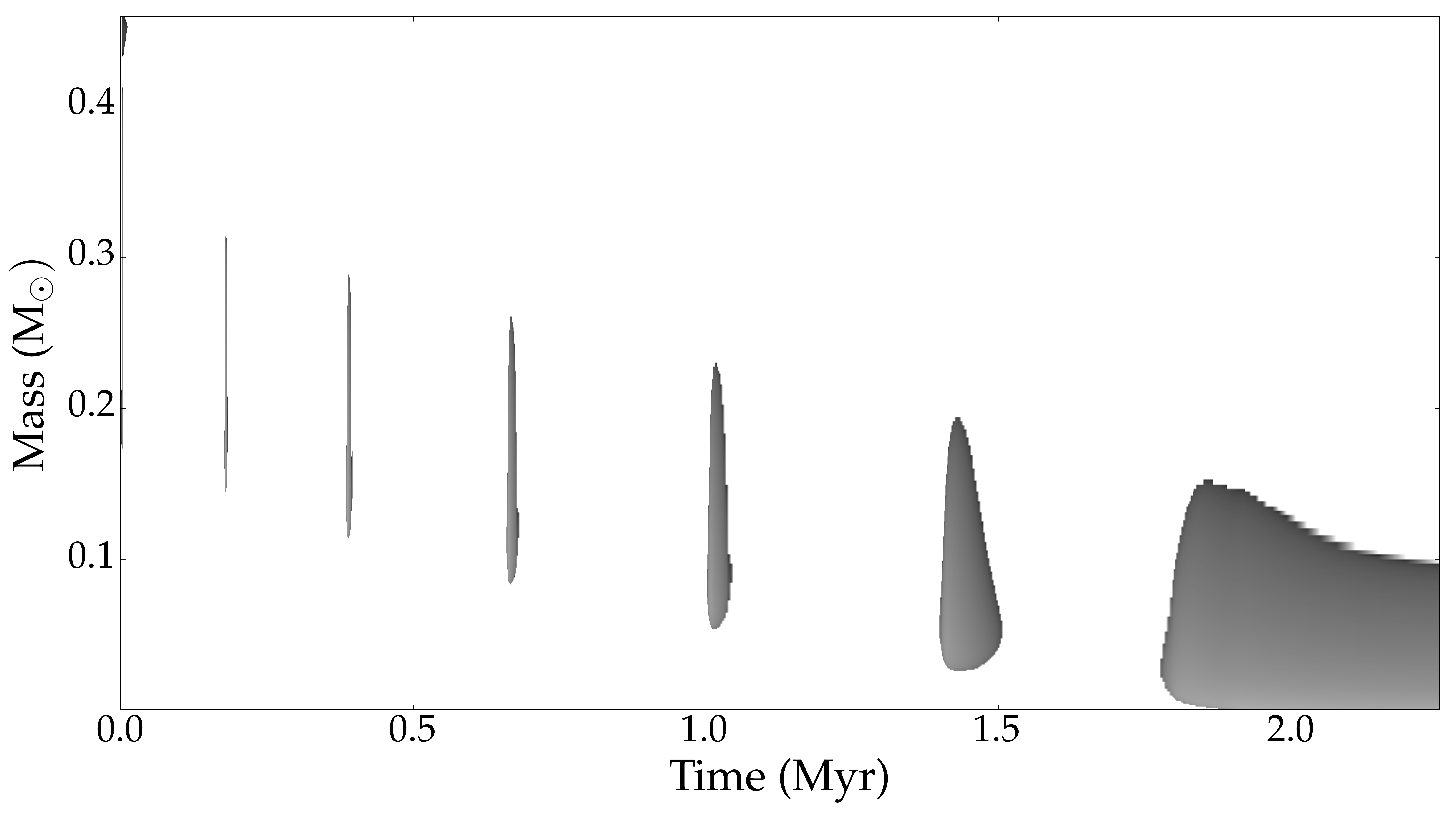}
\caption{Illustration of the helium flash of the  1.5\,M$_\odot$ star. 
Dark regions show convective mixing created by the successive helium sub-flashes. 
White regions show the inert radiative regions.}
\label{fig:kipp1}
\end{figure}

To mimic the effect of binary interaction and envelope mass stripping to form an
sdB star, we use the {\tt relax\_mass} option in MESA, starting at the tip of the red
giant branch (TRGB).  This is achieved by including a huge mass loss rate in
such a way that the degenerate core is surrounded by a very thin inert hydrogen
envelope of mass 0.0008 $M_{\odot}$ after the envelope mass stripping.  In the
next step, the model was evolved to the onset of the successive core helium
flashes.  The weak interaction of neutrinos with matter leads to effective
cooling of the central regions of the degenerate helium core.  During the helium
flashes almost five percent of helium gets burnt into carbon and oxygen.  The
energy released by the nuclear processes increases the core temperature and
consequently the de Broglie wavelength of the electrons, removing the electron
degeneracy from the helium core.  Fig.\,\ref{fig:kipp1} shows the Kippenhahn
diagram illustrating the inward progression of the helium flashes in
$\sim$2\,Myr \citep[see also][]{2012ApJ...744L...6B}. Finally, stable helium
burning through the triple alpha reaction produces carbon during the EHB phase.

The convective core evolution during the EHB is not homogeneous, but rather a
progression through various states with almost unique characteristics.  In
total, there are three stages that can be distinguished from each other. The
first stage is the period between the Zero Age Extreme Horizontal Branch (ZAEHB)
and the time when almost half of the helium is burnt.  During this stage, the
convective core grows uniformly.  The second stage continues until about 10
percent of helium in core remain, while the third stage covers the final EHB
evolution. During this third stage, so-called breathing pulses \citep[see,
e.g.][for a definition]{1985ApJ...296..204C} occur due to the efficiency of
nuclear conversion of carbon to oxygen, which leads to a higher opacity.  Hence,
the convective core grows rapidly and mixes large amounts of helium into the
burning region.  The number of pulses is very sensitive to the amount of
overshooting.  If the convective core boundary reaches the outer layers with low
enough temperature, the pulses will occur prior to the nuclear fuel exhaustion
in the core.

\begin{figure}
\includegraphics[width=\columnwidth]{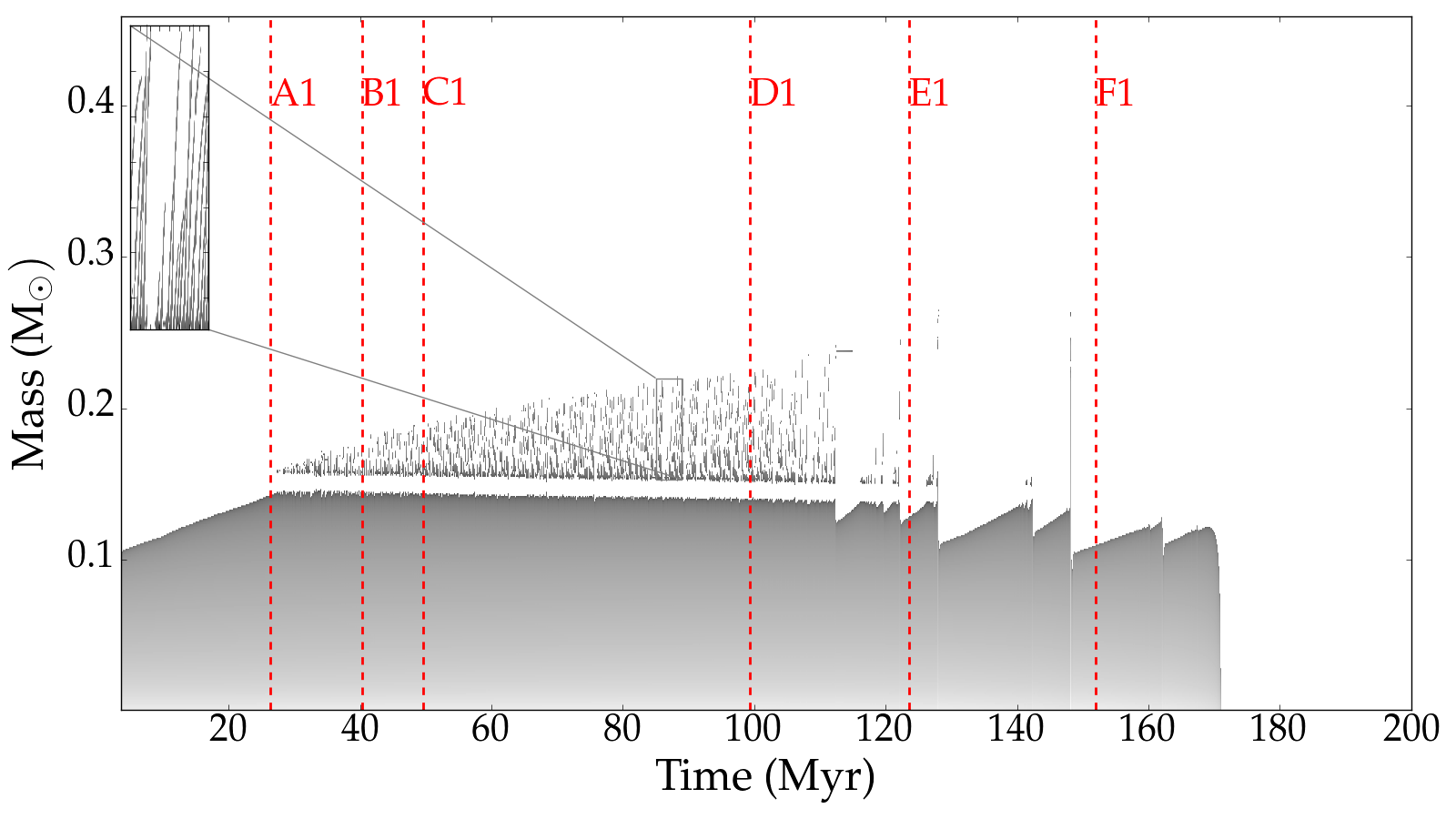}
\caption{Evolution of the He burning core (dark regions) during the EHB phase,
 For the model with moderate overshooting described by the value $f=0.01$.
Temporal appearances and disappearances of convective shells on top of the core occur, 
with narrow inert radiative regions (white) in between.}
\label{fig:kipp(moderate)}
\end{figure}

\begin{figure}
\includegraphics[width=\columnwidth]{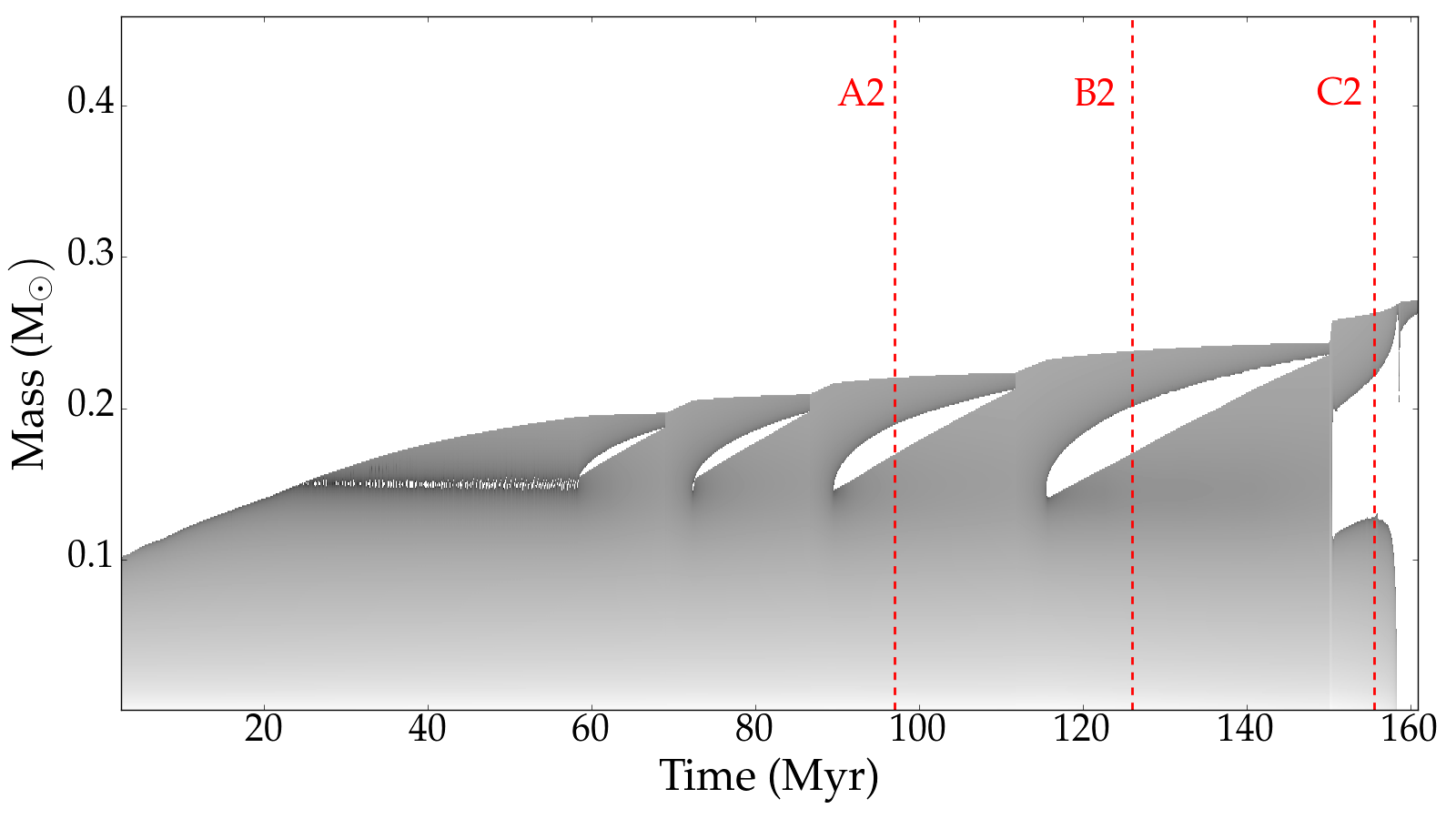}
\caption{Evolution of the He burning core (dark regions) during the EHB phase,
for the model with small overshooting ($f=10^{-5}$). 
The core grows monotonically until $\sim$25 Myr, where it suddenly splits into two convective regions.}
\label{fig:kipp(small)}
\end{figure}

\begin{figure}
\includegraphics[width=\columnwidth]{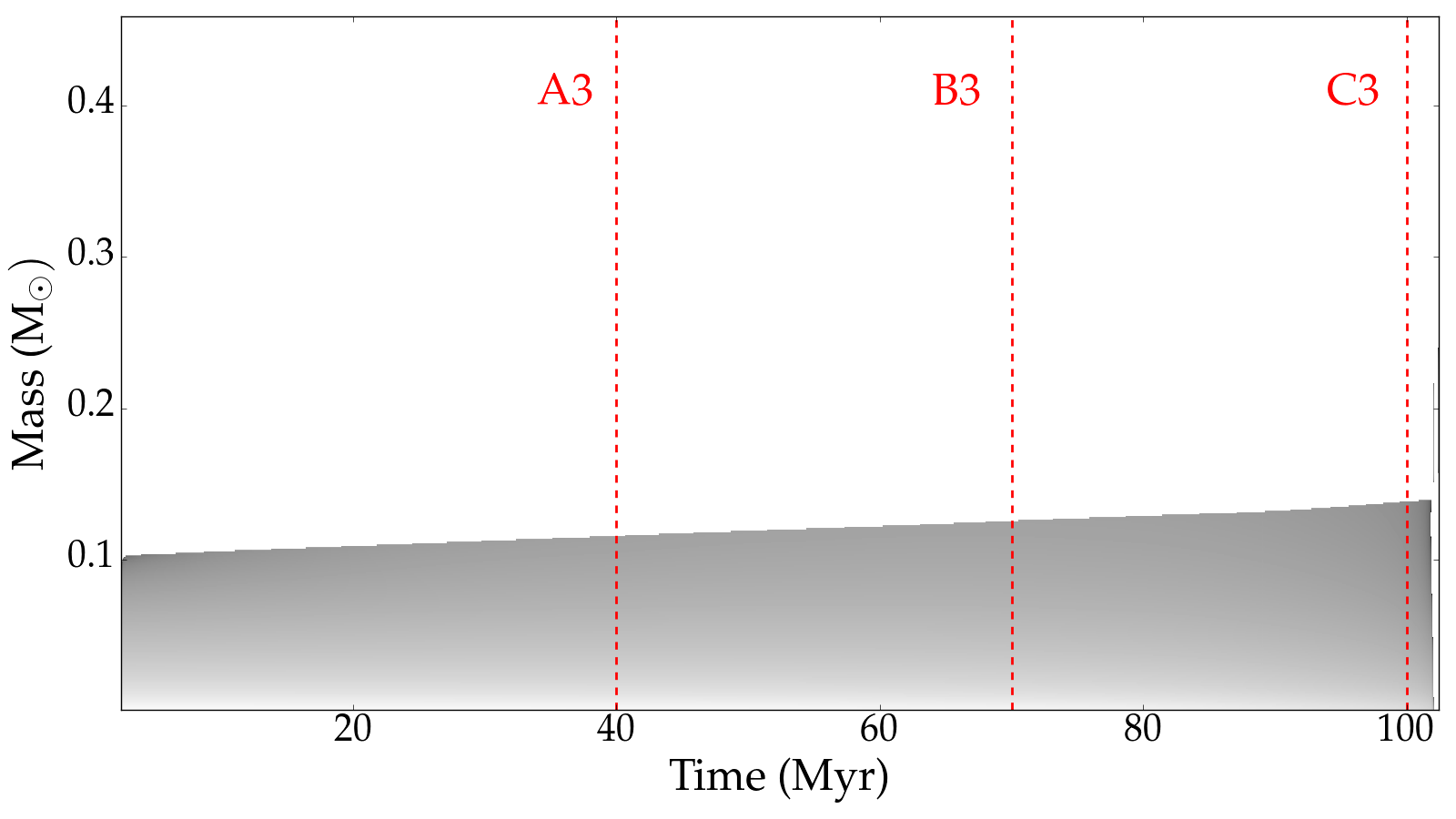}
\caption{Evolution of the He burning core (dark regions) during the EHB phase,
for the model with very small overshooting ($f=10^{-6}$). 
The core grows monotonically during EHB phase.}
\label{fig:kipp(very-small)}
\end{figure}

\begin{figure*}
\includegraphics[width=\textwidth]{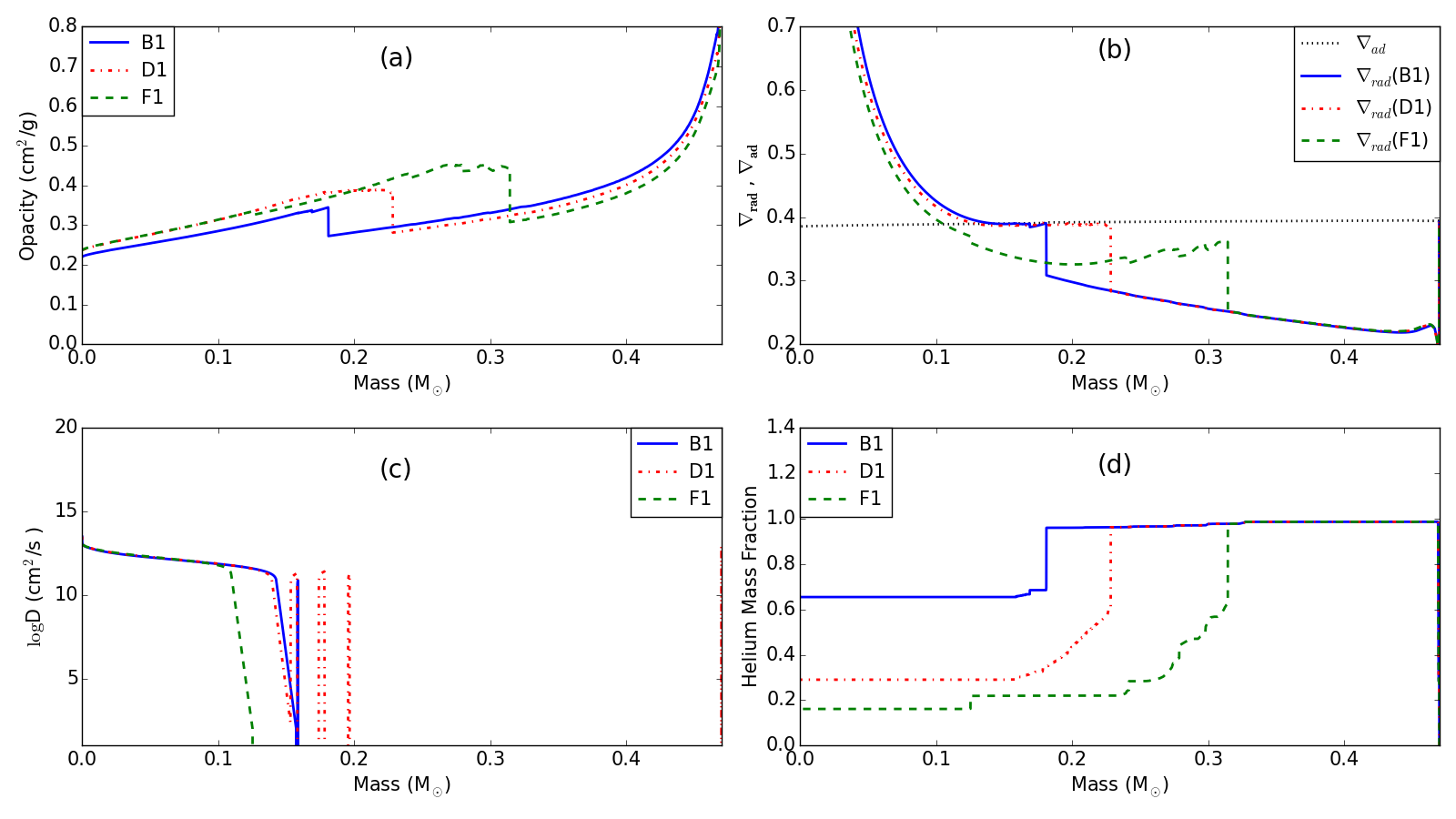}
\caption{Physical quantities profiles of the model with moderate overshooting for 
points B1, D1 and F1 on the Fig.\,~\ref{fig:kipp(moderate)}. (a) Opacity. (b) $\nabla_\mathrm{rad}$ and $\nabla_\mathrm{ad}$. 
(c) Diffusion coefficient. (d) Helium mass fraction }
\label{moderate-evol.png}
\end{figure*}

\begin{figure*}
\includegraphics[width=\textwidth]{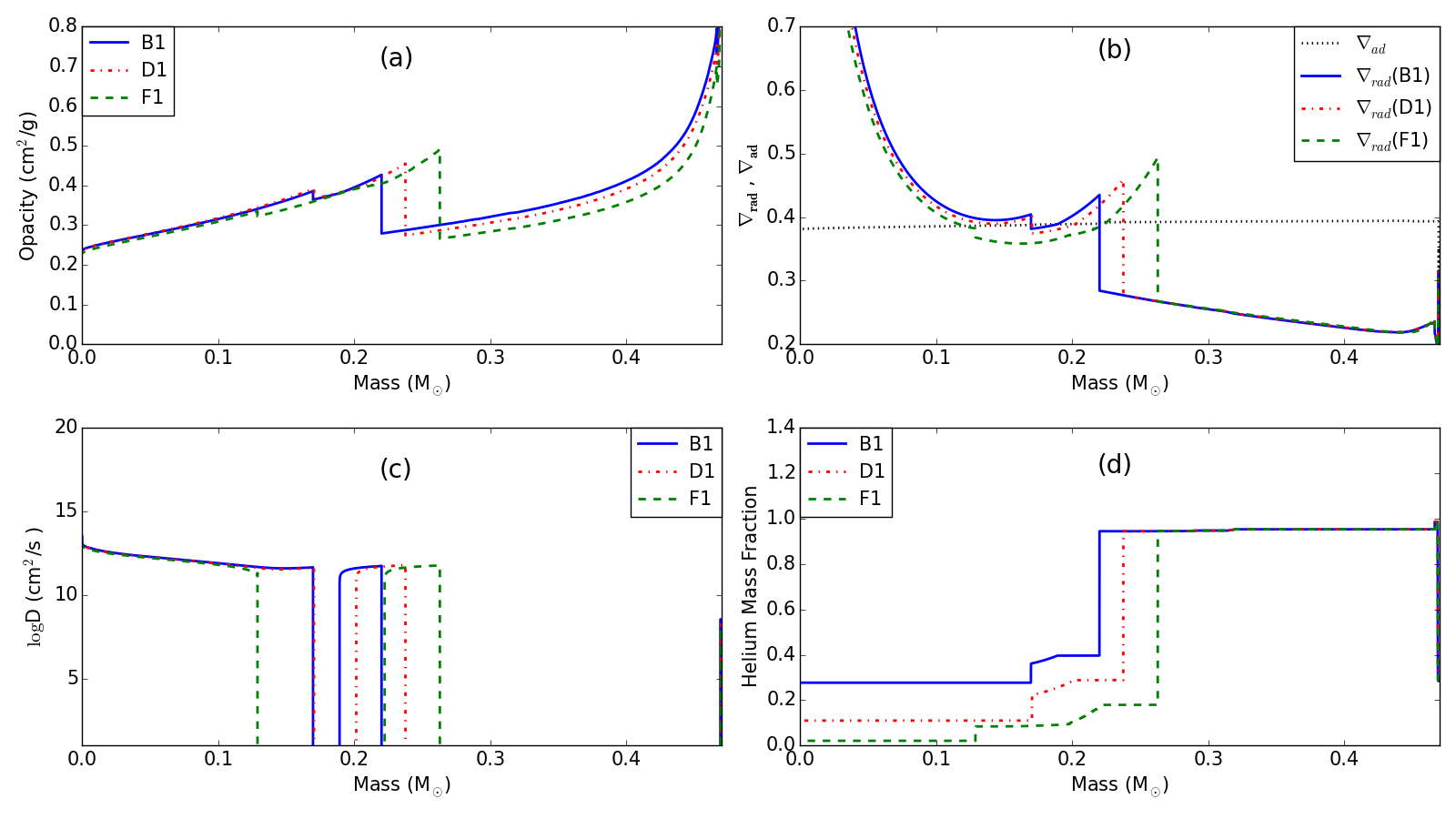}
\caption{Similar to Fig \ref{moderate-evol.png} but for a  model with small overshooting for 
points A2, B2 and C2 on the Fig.\,~\ref{fig:kipp(small)}.}
\label{small-evol.png}
\end{figure*}

\begin{figure*}
\includegraphics[width=\textwidth]{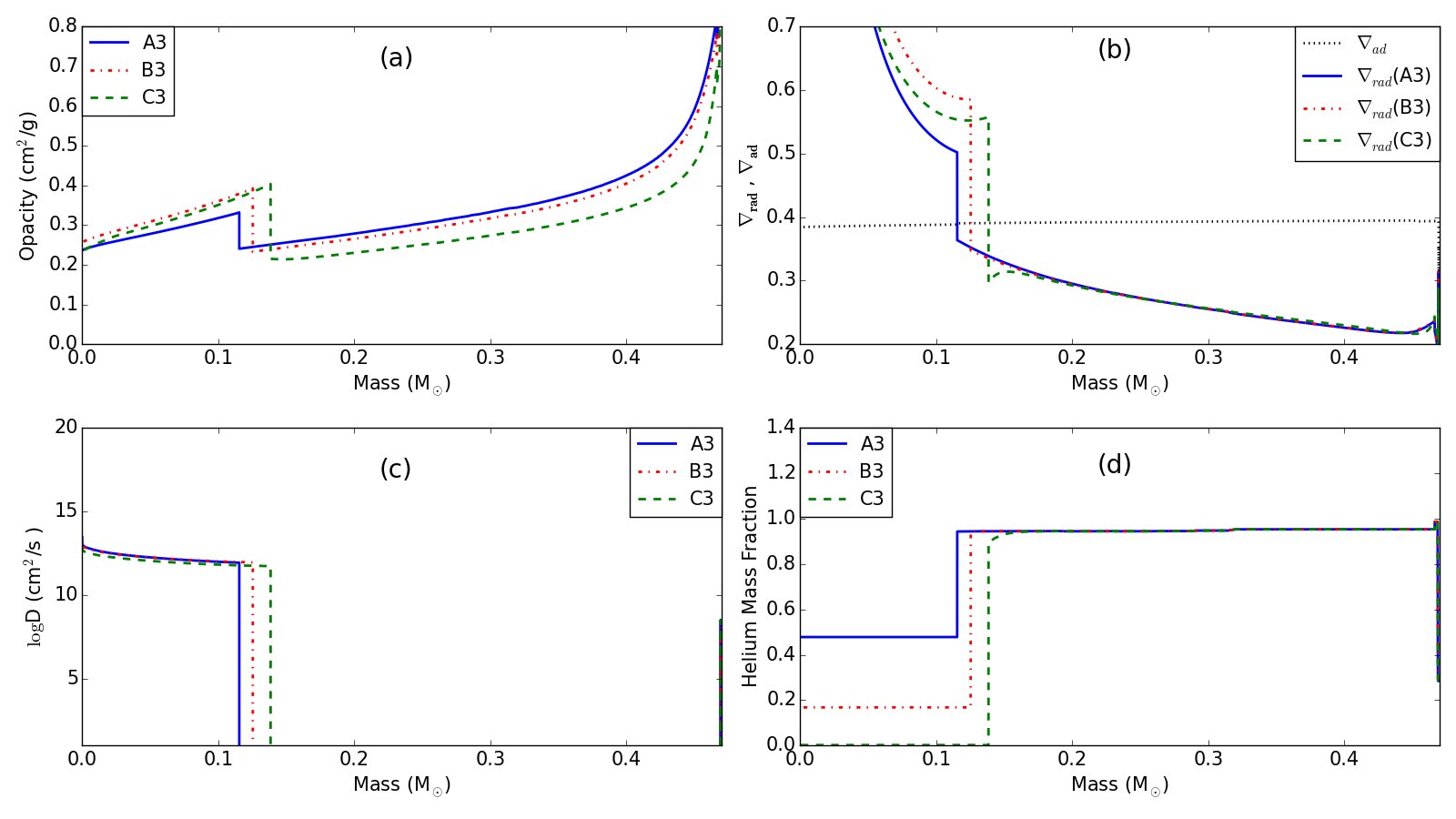}
\caption{Similar to Fig \ref{moderate-evol.png} but for a  model with very small overshooting for 
points A3, B3 and C3 on the Fig.\,~\ref{fig:kipp(very-small)}.}
\label{very-evol.png}
\end{figure*}

During the EHB, models based on three overshooting scenarios were considered:

\begin{itemize}

  \item Moderate overshooting with $f=0.01$, which is shown in
  Fig.\,\ref{fig:kipp(moderate)}. The transfer of carbon and oxygen into the radiative part 
  increases the opacities and hence the convective core grows progressively at first 
  (see, e.g point A1 in Fig.\,\ref{fig:kipp(moderate)}). 
  After first stage, in addition to the inner convective core, outer convective shells occur 
  (see, e.g points B1, C1, D1, E1, F1 in Fig.\,\ref{fig:kipp(moderate)}). 
  At the inner part of the convective core, the burning power
  generated by the triple alpha nuclear fusion is very temperature dependent.
  As a consequence, in this region convection is the dominant source of energy transport.

  At the outer region of the convective core, on one hand, the transfer of
  carbon and oxygen into the radiative part increases the opacities. On the
  other hand, follows Kramers' law, i.e., $\kappa \propto \rho T^{-3.5}$, the
  temperature decreases towards the outer part of core (see
  Fig.\,\ref{moderate-evol.png}(a)) and hence the opacity decreases. Hence, the
  regions with lower $\nabla_{\rm rad}$ near the core become radiative and a
  convective shells occurs simultaneously at the outer regions (see
  Fig.\,\ref{moderate-evol.png}(b)).  Because of mixing due to the overshooting
  between the convective shells and the outer He-rich radiative zone (see
  Fig.\,\ref{moderate-evol.png}(c)), fresh helium can enter the shells, lowering
  the opacity in the shells locally; consequently, the inner base of the
  convective shells become radiative, while on the other hand, carbon and oxygen
  can enter the outer He rich radiative zones, increase the opacity locally;
  hence, the radiative zones near the outer base of convective shells become
  convective.  Convective shells move along the outer regions and finally
  disappear.

  \item Small overshooting with $f = 10^{-5}$, which is shown in
  Fig.\,\ref{fig:kipp(small)}. The convective core grows uniformly at first, and then the convective
  region splits into an inner convective core and an outer convective shell (see, e.g point A2, B2 in Fig.\,\ref{fig:kipp(small)}). 
  Due to the transport of less opaque matter into the convective region,
  the regions inside the convective core become radiative and a convective shell remains
  at the outer regions (see Fig.\,\ref{small-evol.png}(a),(b)). Because of the mixing due to the overshooting at the external
  border of the convective shell, the entire convective shell stabilizes, and becomes
  radiative. Several abrupt breathing pulses occur at the third stage of the core
   He burning (see, e.g point C2 in Fig.\,\ref{fig:kipp(small)}). 
   Similar He burning models have already been calculated for the convective
  core by, e.g. \citet{1990ASPC...11....1S}.

  \item Very small overshooting with $f = 10^{-6}$, which is shown in 
  Fig.\,\ref{fig:kipp(very-small)}. The convective core grows monotonically 
  (see, e.g point A3, B3, C3 in Fig.\,\ref{fig:kipp(very-small)}) and reaches its
  maximum size towards the end of core helium burning.  Carbon and oxygen are
  more opaque than helium. Thus, an excess in the local abundance of carbon and
  oxygen increases the local Rosseland mean opacity $\kappa_R$ (see Fig.\,\ref{very-evol.png}(a)), and consequently
  increases the radiative temperature gradient $\nabla_{\rm rad}\propto\kappa_R$ (see Fig.\,\ref{very-evol.png}(b)).
  As a result, the convective core size grows steadily for all stages of the helium burning. 
\end{itemize}

\section{Asteroseismic aspects}\label{s-astero}

Asteroseismology is an astrophysical tool to unravel the internal structure of stars, due to the
strong dependence of oscilation frequencies on the properties of the stellar interior 
\citep{1989nos..book.....U,2010aste.book.....A}.
Since p-modes are most sensitive to the structure of the outer layers while g-modes 
to the deep interior,  density and chemical stratification near the edge of a convective core have large 
effects on the g-mode behaviour.

\subsection{Period spacings}

Asymptotic analysis predicts equally spaced periods for high-order g-modes.
On the other hand, the measured period spacings of g-modes are known to deviate from
uniformity. Here, we exploit such deviations in terms of the properties of the
stellar interior, more particularly the character of the mixing.

The asymptotic analysis of the periods $p_{n}$ of low-degree $l$, high-order $n$
g-modes of a non-rotating non-magnetic stellar model with homogeneous chemical composition
as derived by \cite{1980ApJS...43..469T} results in:

\begin{equation}
p_{n} =\frac{\pi ^{2}}{\sqrt{l(l+1)} \int_{x_{0}}^{1}\frac{\left | N \right |}{x}dx} (2n + n_{e}),
\end{equation}
where $n_{e}$ is polytropic index of the the surface layer and is a constant
offset for each $l$ and $x$ is the normalised radius.
The Brunt-V\"ais\"al\"a frequency $N^{2}$ is defined as

\begin{equation}    
N^{2} = \frac{g}{r}\left[ \frac{1}{\Gamma_{1}} \frac{d \ln P }{d \ln r} - \frac{d \ln \rho }{d \ln r} \right],
\end{equation}
with g the local gravitational acceleration and $\Gamma_{1} = ({\partial \ln P}  /  {\partial \ln \rho})_{s}$ . 
$N^{2}$ can be rewritten into a more applied form

\begin{equation}    
N^{2} = \frac{g^{2}\rho }{p}\frac{\chi_{T}}{\chi_{\rho}}  (\bigtriangledown_{ad} - \bigtriangledown  + B ),
\end{equation}
with $\nabla = ({\partial \ln T}  /  {\partial \ln p})$ the temperature gradient, $\nabla_\mathrm{ad}$ the adiabatic
temperature gradient, $\chi_{\rho} = ({\partial \ln P}  /  {\partial \ln \rho})_{T}$ 
and $\chi_{T} = ({\partial \ln P}  /  {\partial \ln T})_{\rho}$. B is called the Ledoux term, and it includes 
the effect of composition gradients \citep{1989nos..book.....U,1991ApJ...367..601B, 2013ApJS..208....4P}. 
The asymptotic periods are predicted to be equally spaced in the order 
of the mode ($\Delta P_{n}=P_{n+1}- P_{n}=const$). 

\cite{1992ApJS...80..369B,1992ApJS...81..747B}
extended the asymptotic theory to incorporate the effect of a discontinuity in chemical composition and/or the 
presence of an outer convection zone in the white dwarfs. 
\cite{2000ApJS..131..223C} developed further the asymptotic theory to the specific case of sdB stars
with a convective core and a radiative envelope.

Different mixing processes (overshooting, atomic diffusion, rotation, etc.) lead to different
shapes of the chemical gradient outside the convective core and hence the period
spacing pattern depends on the detailed properties of the buoyancy frequency in
the vicinity of convective core \citep{2008MNRAS.386.1487M}. Observed period spacing
patterns can therefore be used to understand and pinpoint what kind of mixing is acting near the
convective core and how efficient the mixing processes are.  

\begin{figure*}
\includegraphics[width=\textwidth]{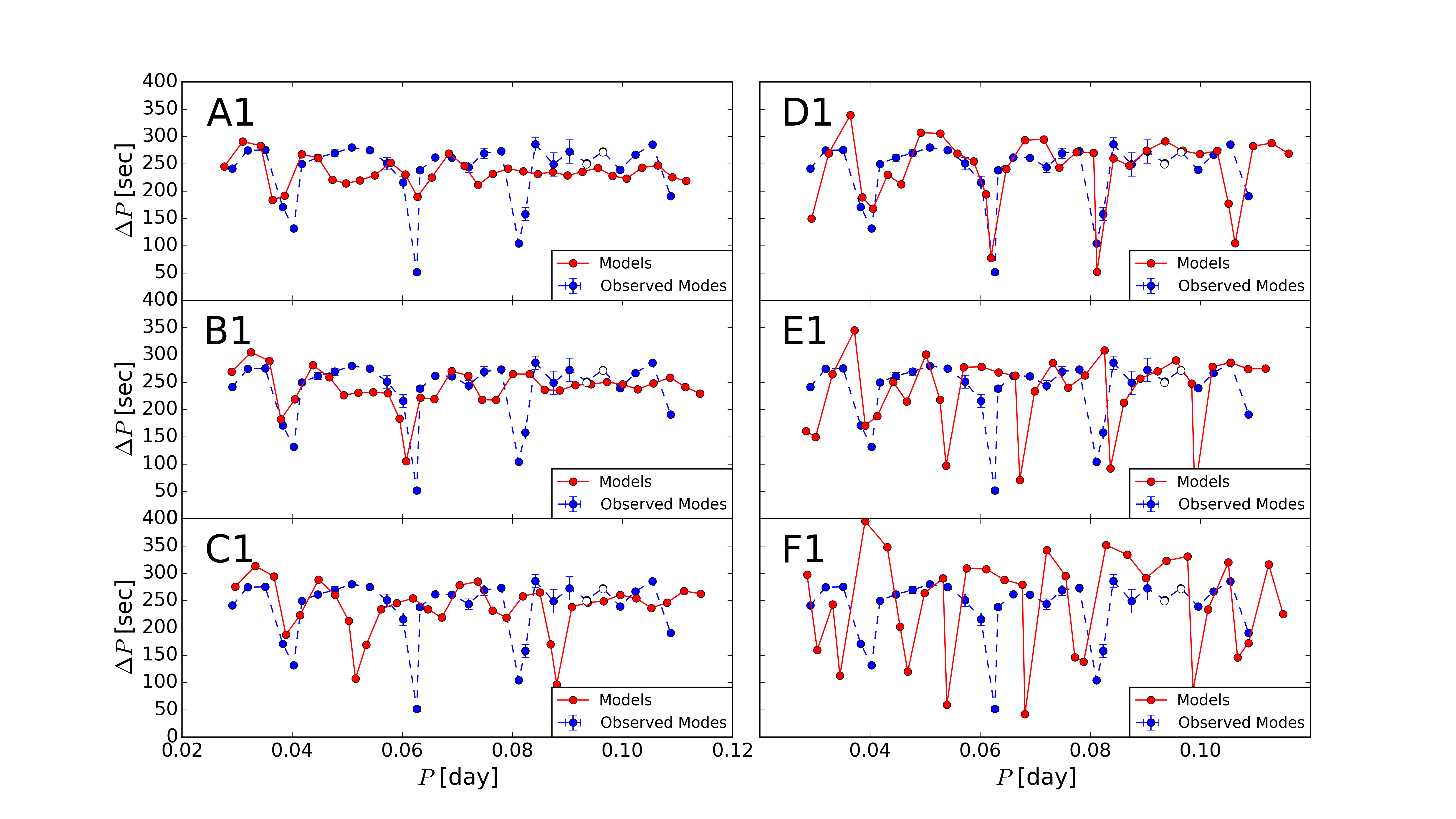}
\caption{Period spacing of the model with moderate overshooting for six points
  indicated in Fig.\,~\ref{fig:kipp(moderate)}}
\label{moderate-seismo.png}
\end{figure*}

\begin{figure}
\includegraphics[width=\columnwidth]{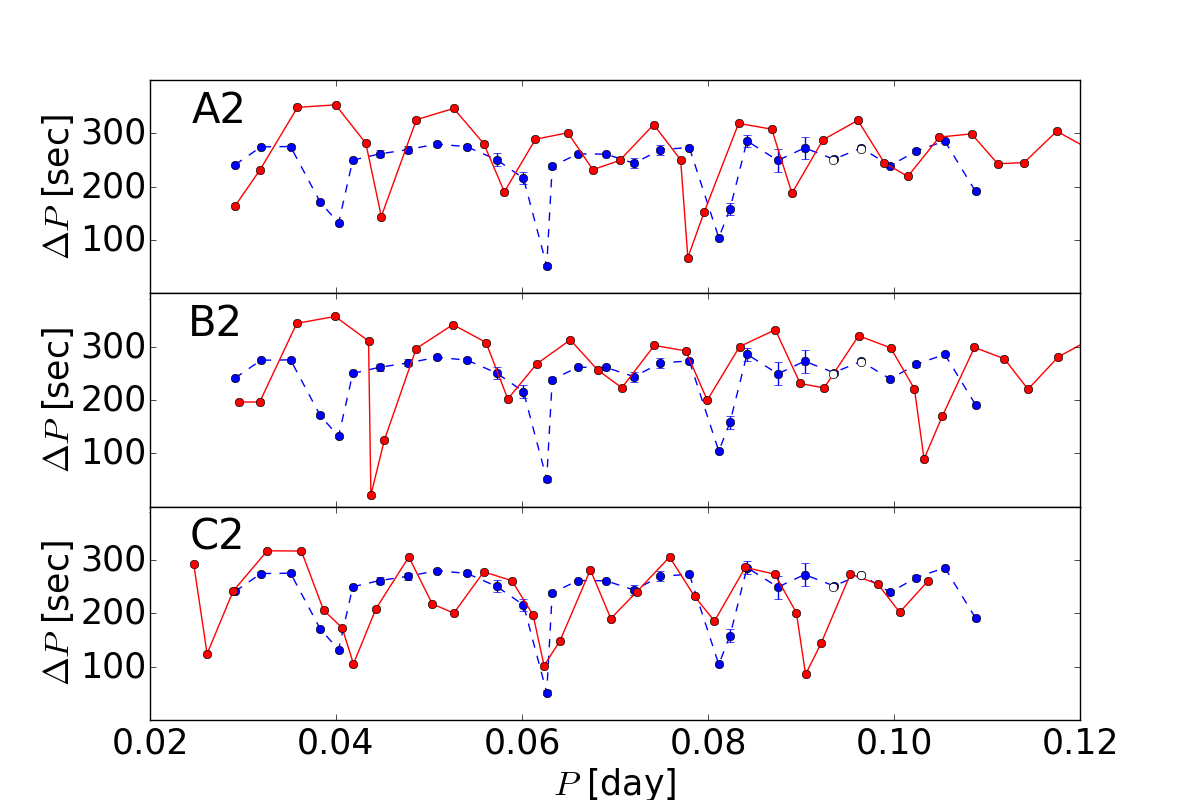}
\caption{Period spacing of the model with small overshooting for three points
  indicated in Fig.\,~\ref{fig:kipp(small)}. }
\label{small-seismo.png}
\end{figure}

For the moderate and small overshooting scenarios, the CO/He transition layer has the most
important influence on the period spacing pattern. Deep mode trapping patterns
can be created by the convective shells. Alongside the evolutionary path, the distance between the sharp composition 
gradients remaining between the convective shells and the lower convective core are larger and the number of 
trapped modes  increases  (Figs.\,\ref{moderate-seismo.png} and \ref{small-seismo.png}).

\begin{figure}
\includegraphics[width=\columnwidth]{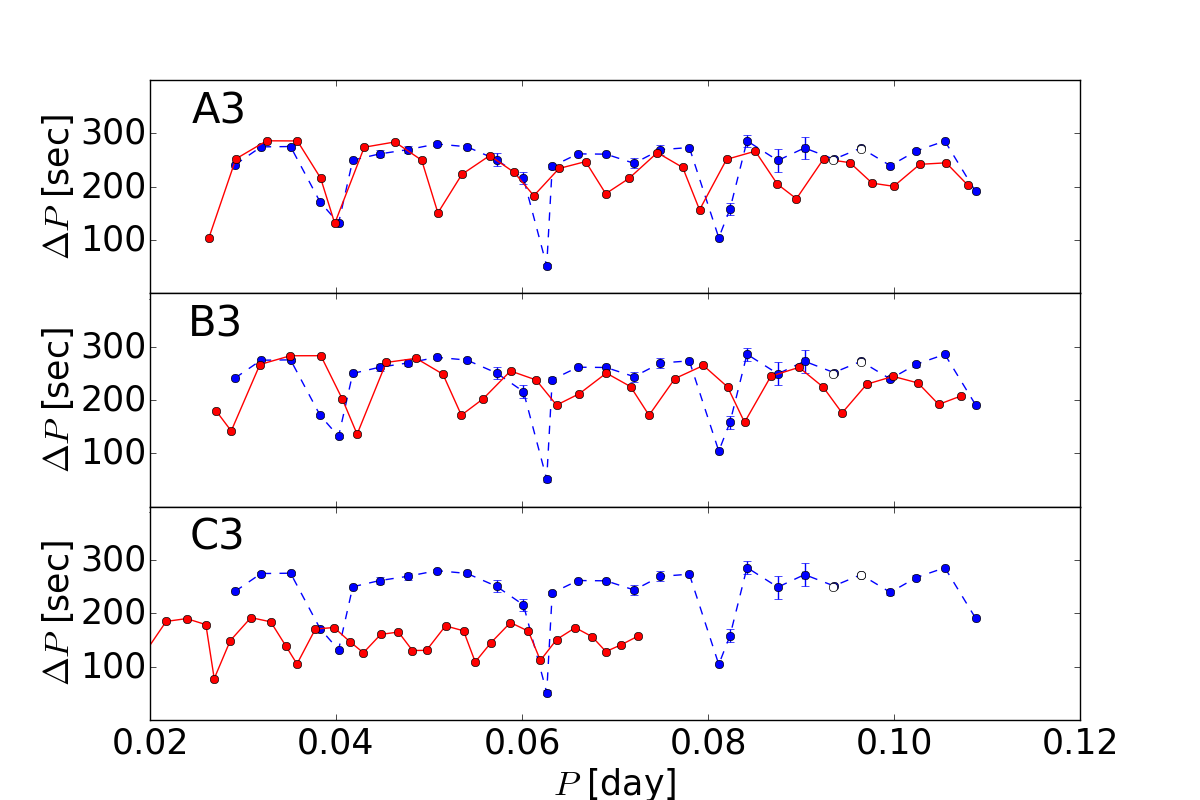}
\caption{Similar to Fig.\,\,\ref{small-seismo.png} but for a model with very
  small overshooting for 
three points indicated in Fig.\,~\ref{fig:kipp(very-small)}.}
\label{very-seismo.png}
\end{figure}

In the very small overshooting scenario, the He/H transition layer has the largest
influence on the period spacing pattern. At the EHB, this He/H transition
interface remains constant. The gradual decrease in the period spacing arises
due to the growing convective core and along with it the shrinking of the
propagation cavity. Therefore, the period spacing pattern depends on the
detailed properties of the buoyancy frequency in the vicinity of He/H transition
layer, cf.\ Fig.\,\ref{very-seismo.png}.
\cite{2000ApJS..131..223C,2002ApJS..139..487C,2002ApJS..140..469C} 
have already demonstrated that the period spacing pattern for such a convective core structure 
is not very sensitive to the presence of the CO/He transition layer.

\subsection{Trapped modes}
 
\begin{figure*}
\begin{center}
$  \begin{array}{cc}
   \includegraphics[width=90mm]{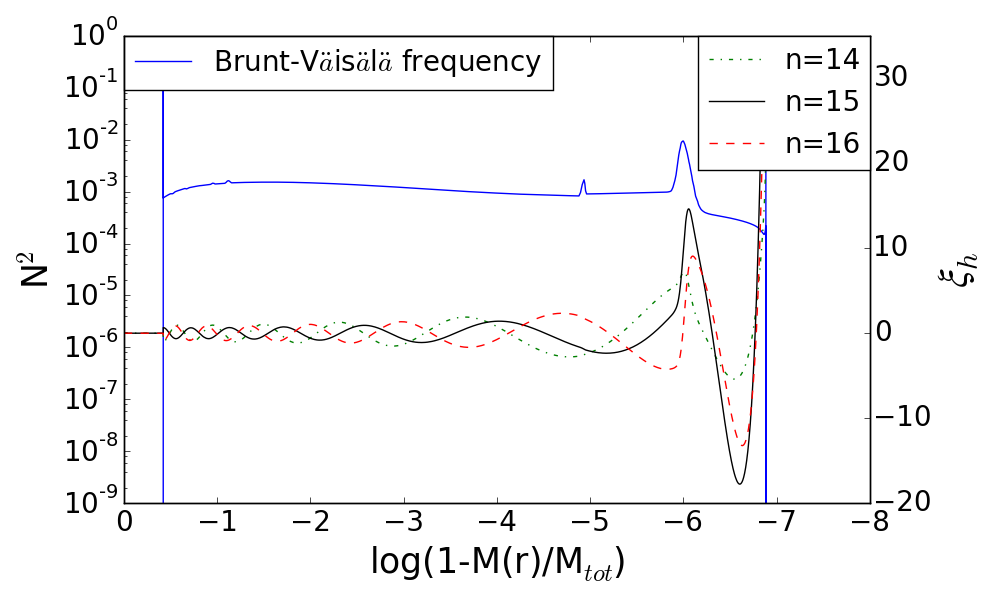}&
   \includegraphics[width=90mm]{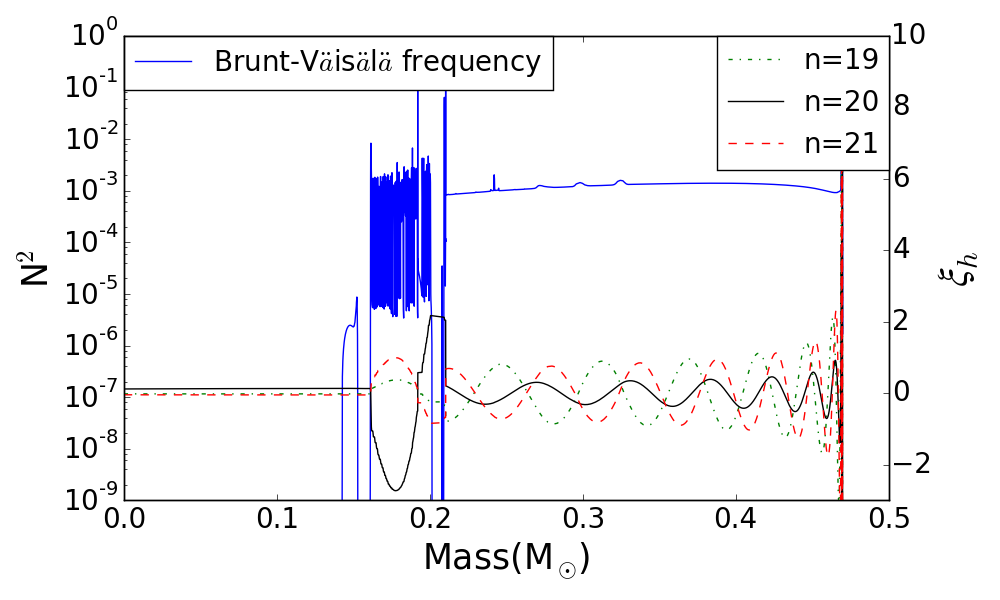}
  \end{array} $
\end{center}
\caption{The the Brunt-V\"ais\"al\"a frequency (left $y$-axis) and the
  horizontal eigendisplacement $\xi_h$ (right $y$-axis) of three dipole g-modes
  with different radial order ($n$).  Right and left panels illustrate the
  behaviour of the trapped modes (black solid lines) near the He/H transition
  layer for models with monotonically growing core based on very small
  overshooting and near the CO/He convective boundary for fluctuating growth
  stages of models with a moderate overshooting.}
\label{pics:bl}
\end{figure*}

The radiative gradient and/or chemical gradients change the density profile and consequently the 
Brunt-V\"ais\"al\"a frequency.  
In sdB models, the H to He transition layer in the outer envelope, and the He to CO transition layer
near the convective boundary implies sharp features in the Brunt-V\"ais\"al\"a frequency.
Such sharp features leave observable fingerprints in the period spacing patterns of g-modes 
\citep{2008MNRAS.386.1487M, 2015ApJ...805..127C}.
The displacement eigenfunctions of trapped modes have two nodes close to the local peak associated 
with the inner transition layer near the core. 
Fig.\,~\ref{pics:bl} shows that the horizontal component of the eigendisplacement $\xi_h$ is trapped in
the vicinity of the Brunt-V\"ais\"al\"a frequency peak (blue solid line). 

The results derived from our models illustrate that the nature of the He/H and
CO/He chemical transition layers determines the effect they have on the mode
trapping.  In the case of mixing due to very small overshooting and the first stage 
of moderate and small overshooting, the CO/He transition
layer has negligible impact on the trapping, and the He/H transition layer
controls the trapping of the modes. The results for this scenario are in good 
agreement with those of \cite{2000ApJS..131..223C,2002ApJS..139..487C,2002ApJS..140..469C}. 
On the other hand, the mixing caused by moderate and small overshooting in fluctuating stages 
reveals that both transition layers play a considerable role on the mode trapping. The He/H transition layer
and the convective shells produce deeply trapped modes.  A sharp density stratification
 at the outer edge of the convective shells change the Buoyancy restoring force and
gives rise to a sinusoidal variation of the amplitude of the eigenfunction.
Some of the displacement eigenfunctions having nodes very close to the
convective shells obtain a large amplitude at the radiative regions between the
convective core and shells (see Fig.\,~\ref{pics:bl}).The results for this scenario are similar to those mentioned in.

\subsection{Mode inertia}

The non-uniform kinetic energy distribution of g-modes resembles the trapping
pattern. Mode inertia is defined as the interior mass influenced by the kinetic energy of any given mode.
The normalized inertia of the modes illustrate the time-averaged
kinetic energy connected with displacement eigenfunction:
\begin{equation}
E=\frac{4\pi\int_{0}^{R} \left [ \left | \tilde{\xi _{r}}(r) \right |^{2} + \sqrt{l(l+1)}\left |
\tilde{\xi _{h}}(r) \right |^{2}  \right ]\rho _{0} r^{2} dr} {M \left [\left | \tilde{\xi _{r}}(R)
\right |^{2} + \sqrt{l(l+1)}\left | \tilde{\xi _{h}}(R) \right |^{2}   \right ]},
\end{equation}
where $\xi _{h}$ and $\xi _{r}$ are the horizontal and radial components of the
displacement eigenfunction.

\begin{figure*}
\begin{center}
  $\begin{array}{cc}
  \includegraphics[width=165mm]{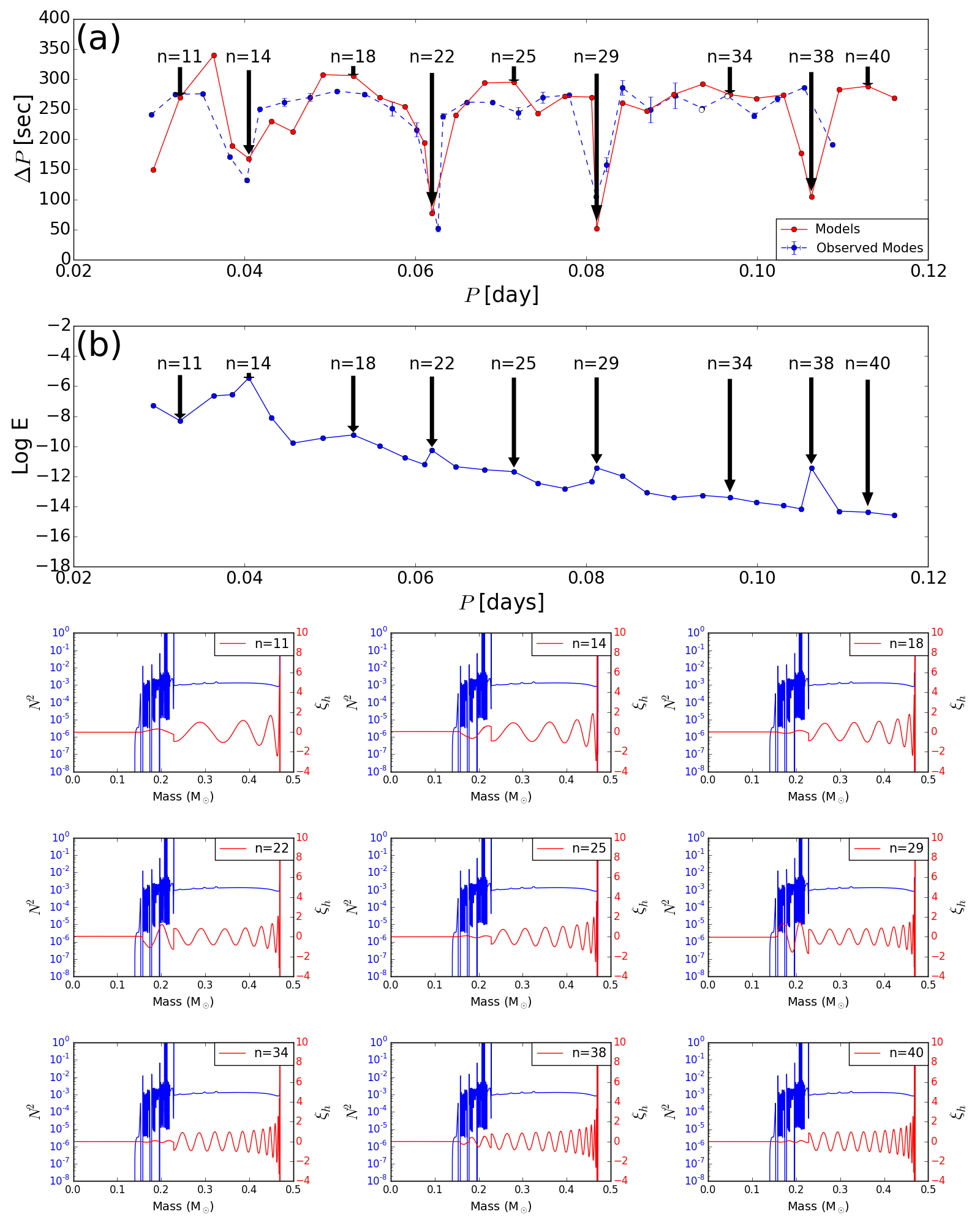}&
  \end{array}$
\end{center}
\caption{Panel a) Period spacing pattern of g-modes with the same degree $(l=1)$
  and consecutive radial order $(n)$ for a model with moderate overshooting for 
  the point D1 on the Fig.\,~\ref{fig:kipp(moderate)}.
  Panel b) The normalized mode inertia as a function of the periods of the dipole g-modes.
  Lower panels: red solid lines show the horizontal displacement eigenfunction
  of the dipole g-modes for different radial order as indicated.  The blue solid
  lines illustrate the Brunt-V\"ais\"al\"a frequency in the stellar interior.}
\label{pics:bl1}
\end{figure*}

\begin{figure*}
\begin{center}$
\begin{array}{cc}
\includegraphics[width=165mm]{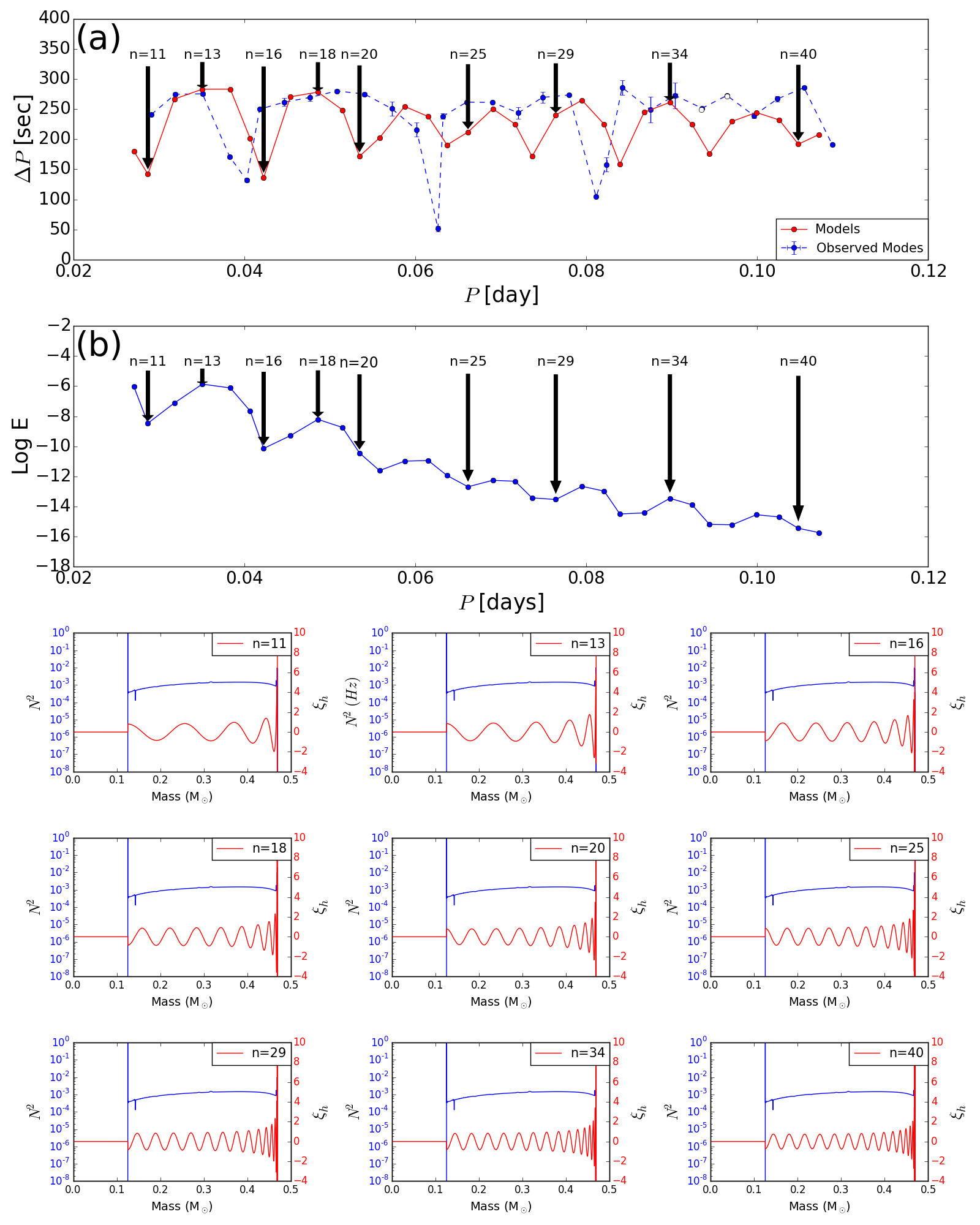}&
\end{array}$
\end{center}
\caption{Same as Fig.\,~\ref{pics:bl1}, for a model with very small overshooting for 
point B2 on the Fig.\,~\ref{fig:kipp(very-small)}.}
\label{pics:bl2}
\end{figure*}

In the fluctuating stages of the scenario with moderate and small overshooting, the few trapped
modes have large amplitude inside the radiative layer between the convective
core and the convective shells. Therefore, such deeply trapped modes have large
mode inertia as can be seen in panel (b) of Fig.\,~\ref{pics:bl1}). Such centrally-enhanced mode
energies connected with high local amplitudes of trapped modes in the deep
stellar interior near the He/CO core boundary have been reported before 
\citep[e.g.][in the case of white dwarfs]{1992ApJS...80..369B,1992ApJS...81..747B}.
For the very small overshooting scenario and uniformly growing stages of other scenarios, 
the trapped modes have lower amplitude at the inner layers of the sdB star in comparison 
with the outer edge of the He/H transition zone. Therefore, such trapped modes have small
mode inertia. They are shown in the panels (b) of Fig.\,~\ref{pics:bl2}.  
\cite{2000ApJS..131..223C,2002ApJS..139..487C,2002ApJS..140..469C} 
have found out that for such a very small overshooting scenario, all g-modes are reflected 
in the vicinity of the convective core boundary. Therefore, the mode inertia is not sensitive 
to the CO/He transition layer.

\section{Diffusion}

Atomic diffusion in MESA includes gravitational settling, thermal and
concentration diffusion.  Gravitational settling leads heavy elements to settle
in the inner regions and concentration diffusion implies charged particles to be
subjected to an electric field. Thermal diffusion causes heavy particles to move
more slowly than light ones at any temperature or kinetic energy.  MESA solves
the Burgers flow equations \citep{1969fecg.book.....B} based on a routine
provided by \citet{1994ApJ...421..828T}, with modifications by 
\citet{2011MNRAS.418..195H} for non Coulomb interactions and radiative
levitation.

\begin{figure*}
\begin{center}
   $\begin{array}{cc}
    \includegraphics[width=90mm]{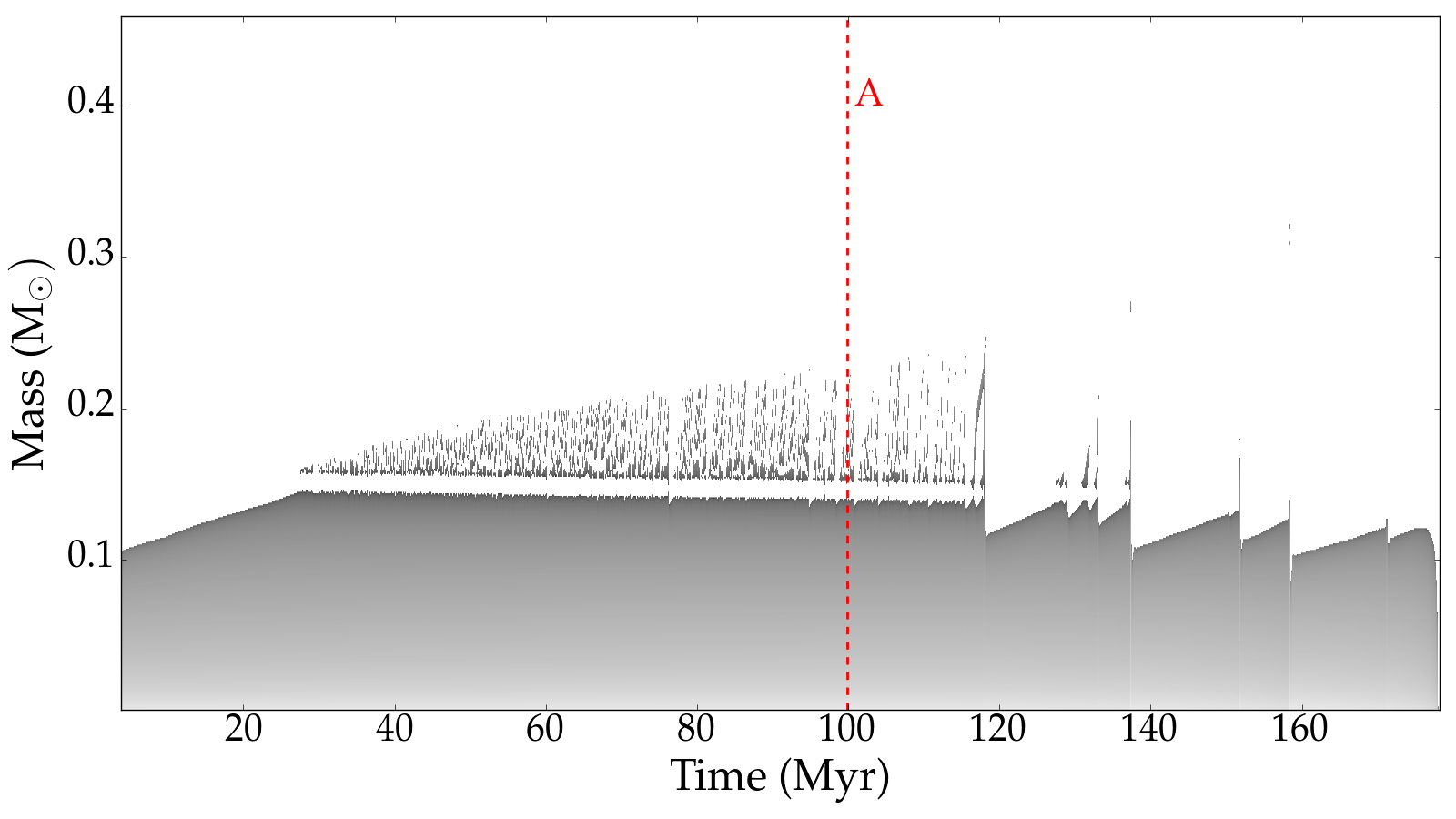}&
    \includegraphics[width=90mm, height=55mm]{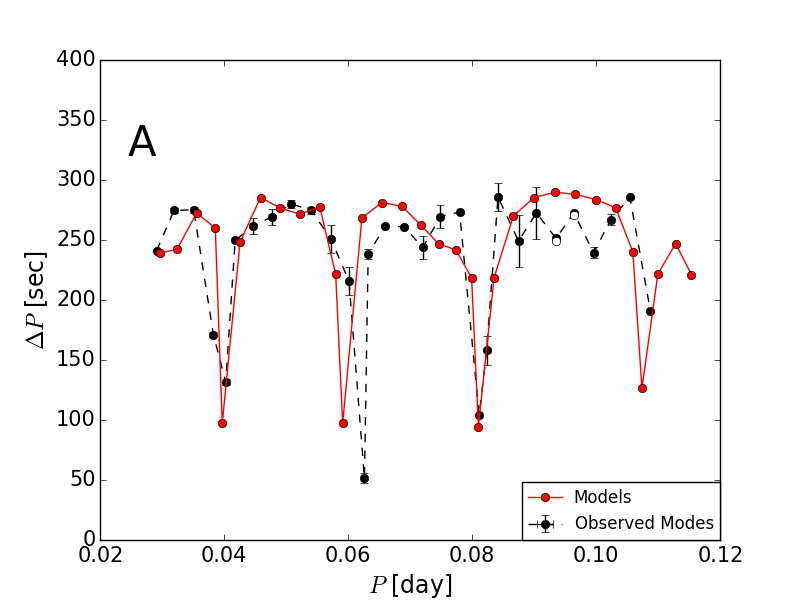}
    \end{array}$
\end{center}
\caption{Evolution of the He burning core (dark regions) during the EHB phase,
 For the model with moderate overshooting described by the value $f=0.01$ in presence of the diffusion.
Temporal appearances and disappearances of convective shells on top of the core occur, 
with narrow inert radiative regions (white) in between.}
\label{kipp-mod-diff.png}
\end{figure*}

\begin{figure*}
\begin{center}
$\begin{array}{cc}
\includegraphics[width=90mm]{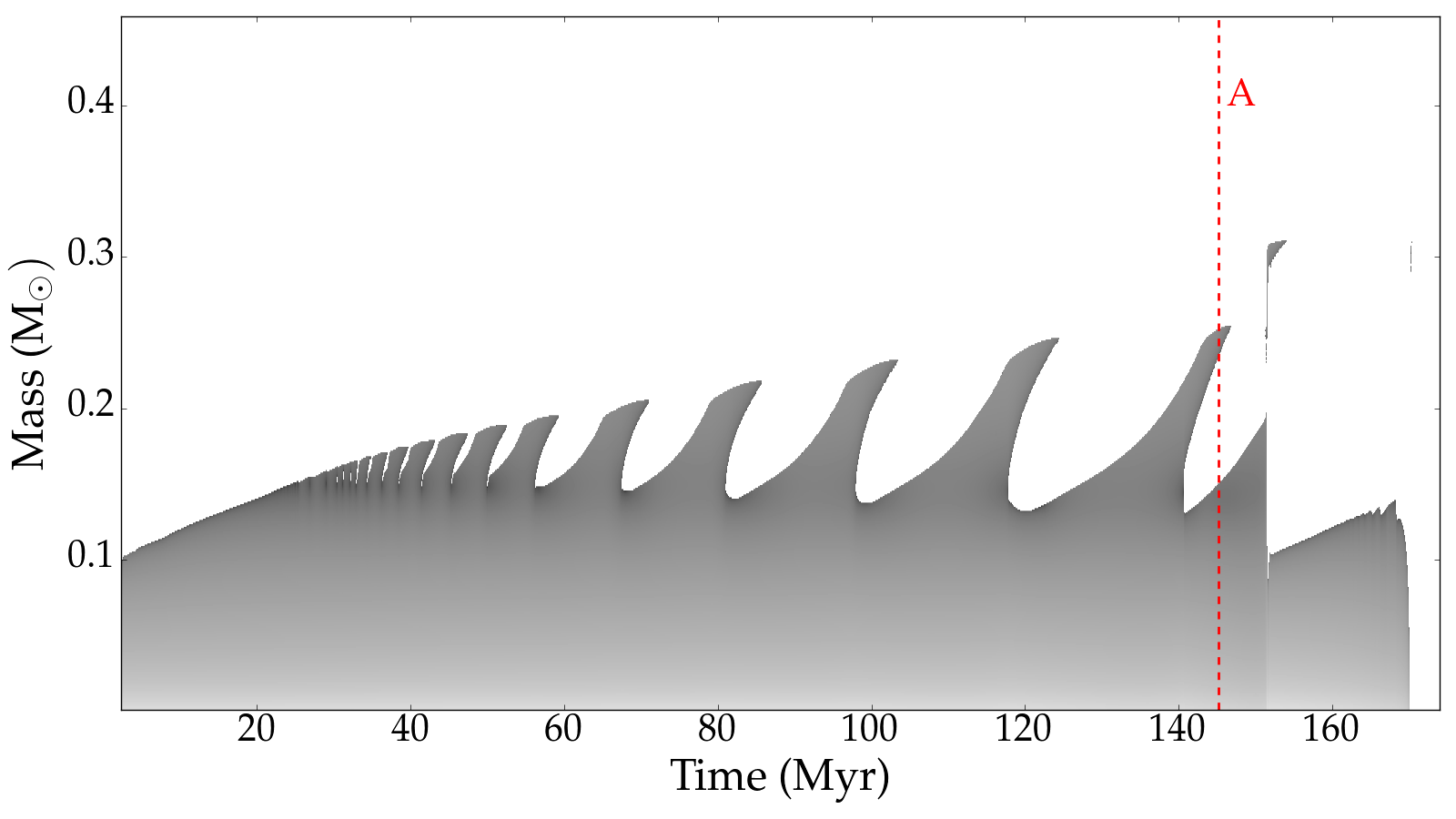}&
\includegraphics[width=90mm, height=55mm]{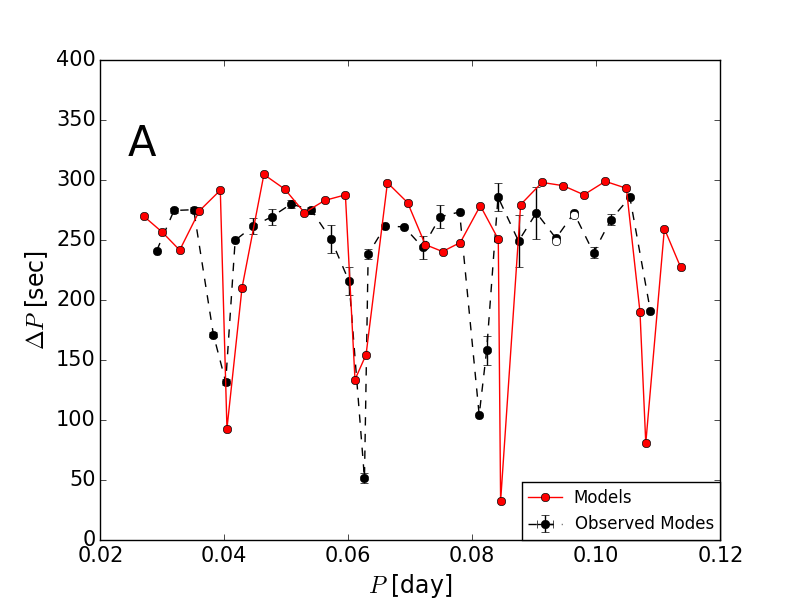}
\end{array}$
\end{center}
\caption{Evolution of the He burning core (dark regions) during the EHB phase,
 For the model with small overshooting described by the value $f=10^{-5}$ in presence of the diffusion.
the convective region splits into an inner convective core and an outer convective shell.}
\label{kipp-small-diff.png}
\end{figure*}

\begin{figure*}
\begin{center}
$\begin{array}{cc}
\includegraphics[width=90mm]{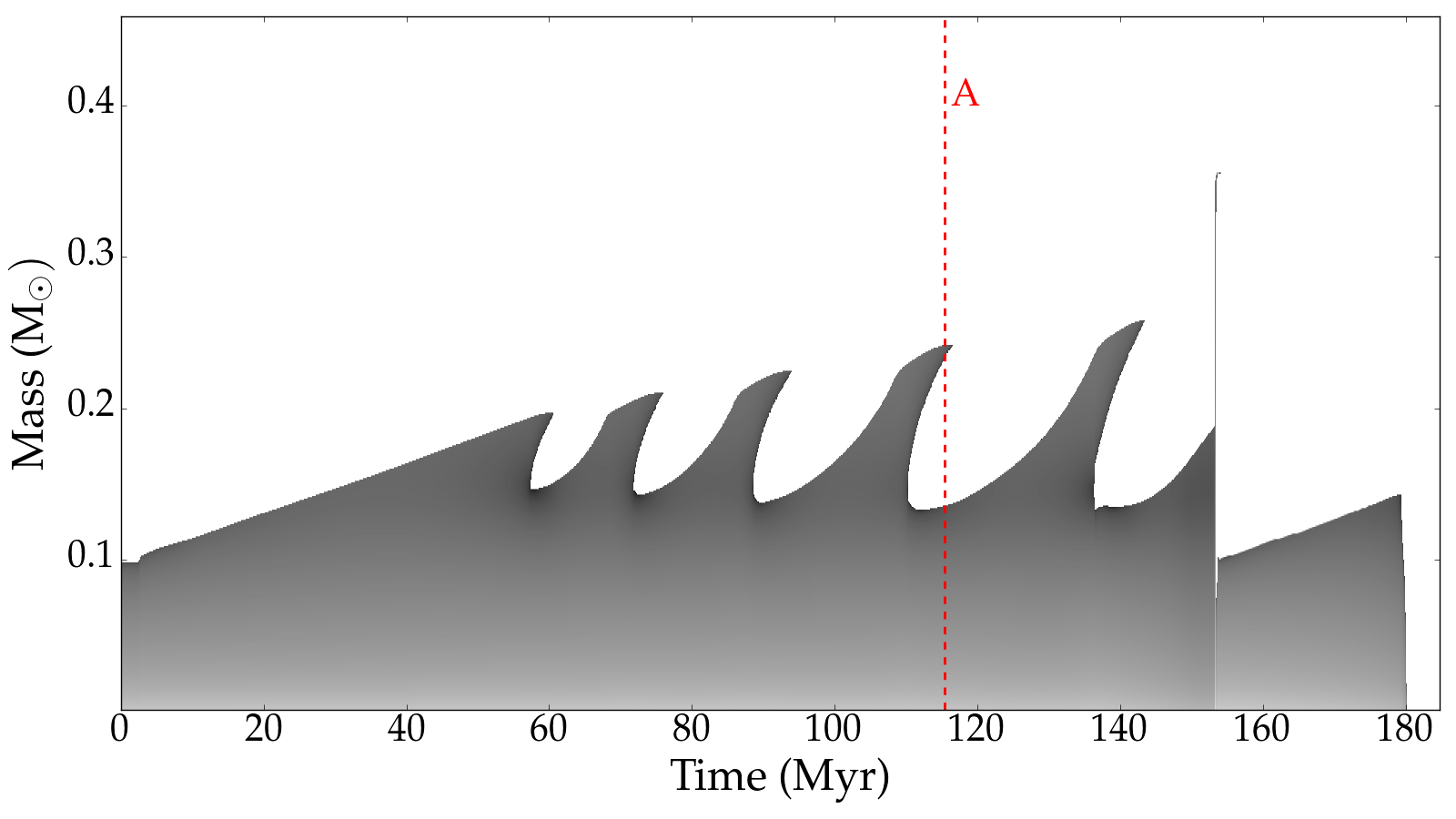}&
\includegraphics[width=90mm, height=55mm]{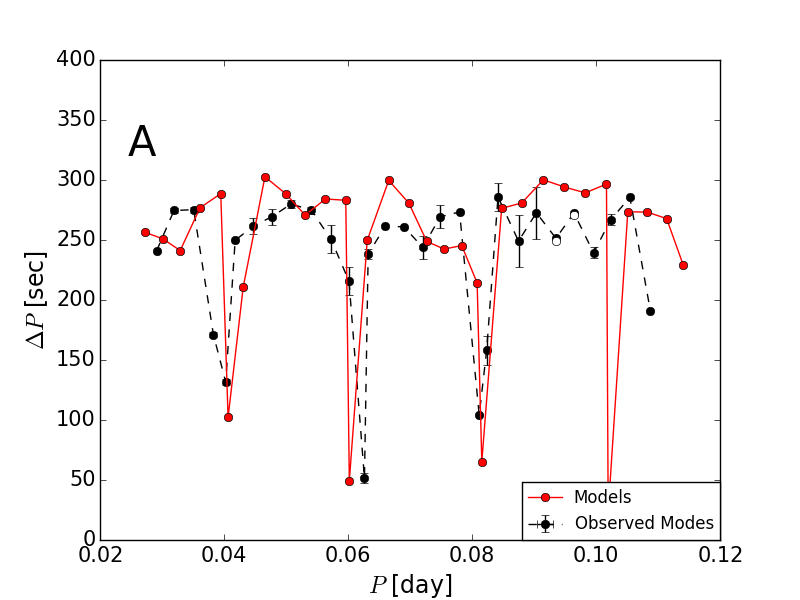}
\end{array}$
\end{center}
\caption{Evolution of the He burning core (dark regions) during the EHB phase,
for the model with very small overshooting $f=10^{-6}$ in presence of the diffusion.}
\label{kipp-very-diff.png}
\end{figure*}

Considering atomic diffusion as the only mixing process between the convective
core and the radiative regions, increases the convective core mass
\citep{2015ApJ...806..178S}.  When moderate or small overshoot is added to the
atomic diffusion, we see the same behaviour of the convective core discussed
above for the moderate or small overshoot models without atomic diffusion. In
these cases, overshooting is the dominant mechanism for the matter transfer
between the convective core and the radiative regions.  The maximum convective
core size and the lifetime are both almost the same (see
Fig.\,~\ref{kipp-mod-diff.png}, ~\ref{kipp-small-diff.png}).

For the very small overshooting models, diffusion leads to a split of the
convective core. Also, diffusion increases the lifetime of the EHB phase (see
Fig.\,~\ref{kipp-very-diff.png}).  Atomic diffusion is also able to change the
profile of the Brunt-V\"ais\"al\"a frequency in the outer layers. Diffusion
broadens the steep composition gradients leading to less efficient mode trapping
in the H-He transition layer. In phases with a monotonically growing core, the
region above the He/H transition layer has the largest influence on the period
spacing pattern because the He/H transition interface is smoothed by atomic
diffusion. On the other hand, at the late phases of models with a convective
shell, the C-O/He transition layer has the most important influence on the
period spacing pattern, due to deep mode trapping patterns created by the
convective shell.

\begin{figure*}
\centering
\begin{tabular}{c c}
\subfloat[]{\includegraphics[width=80mm]{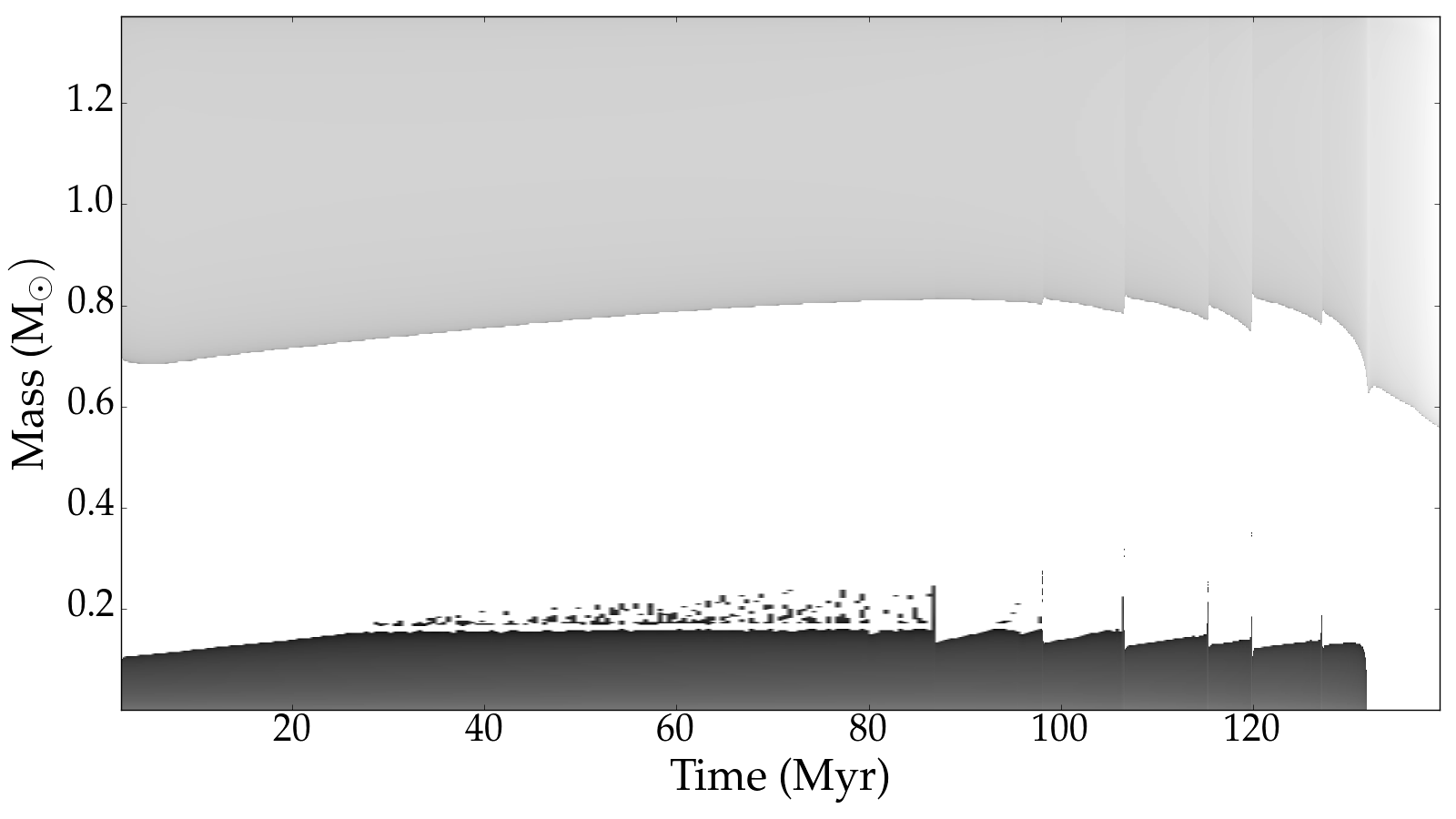}} & 
\subfloat[]{\includegraphics[width=80mm]{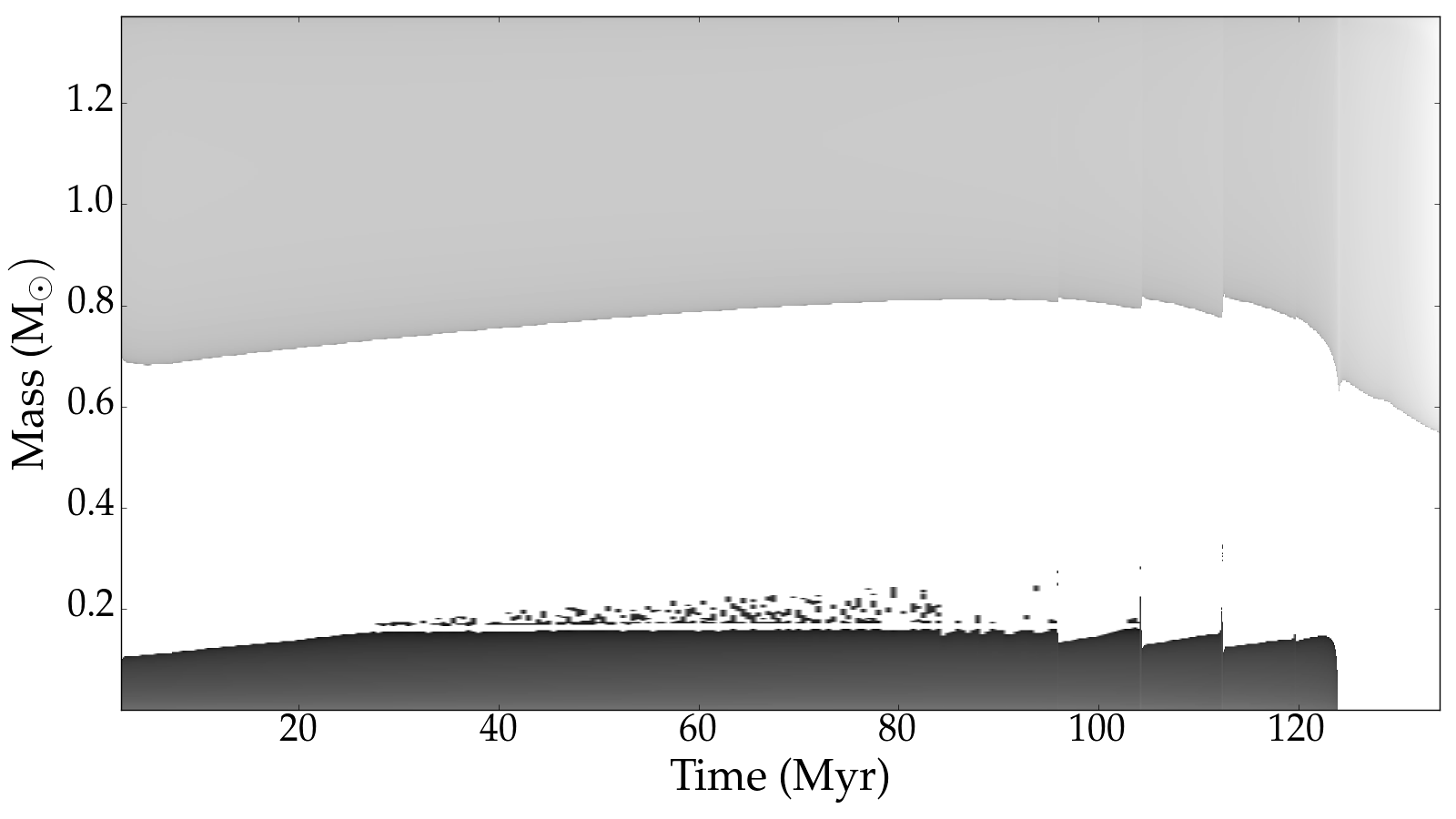}} \\
\subfloat[]{\includegraphics[width=80mm]{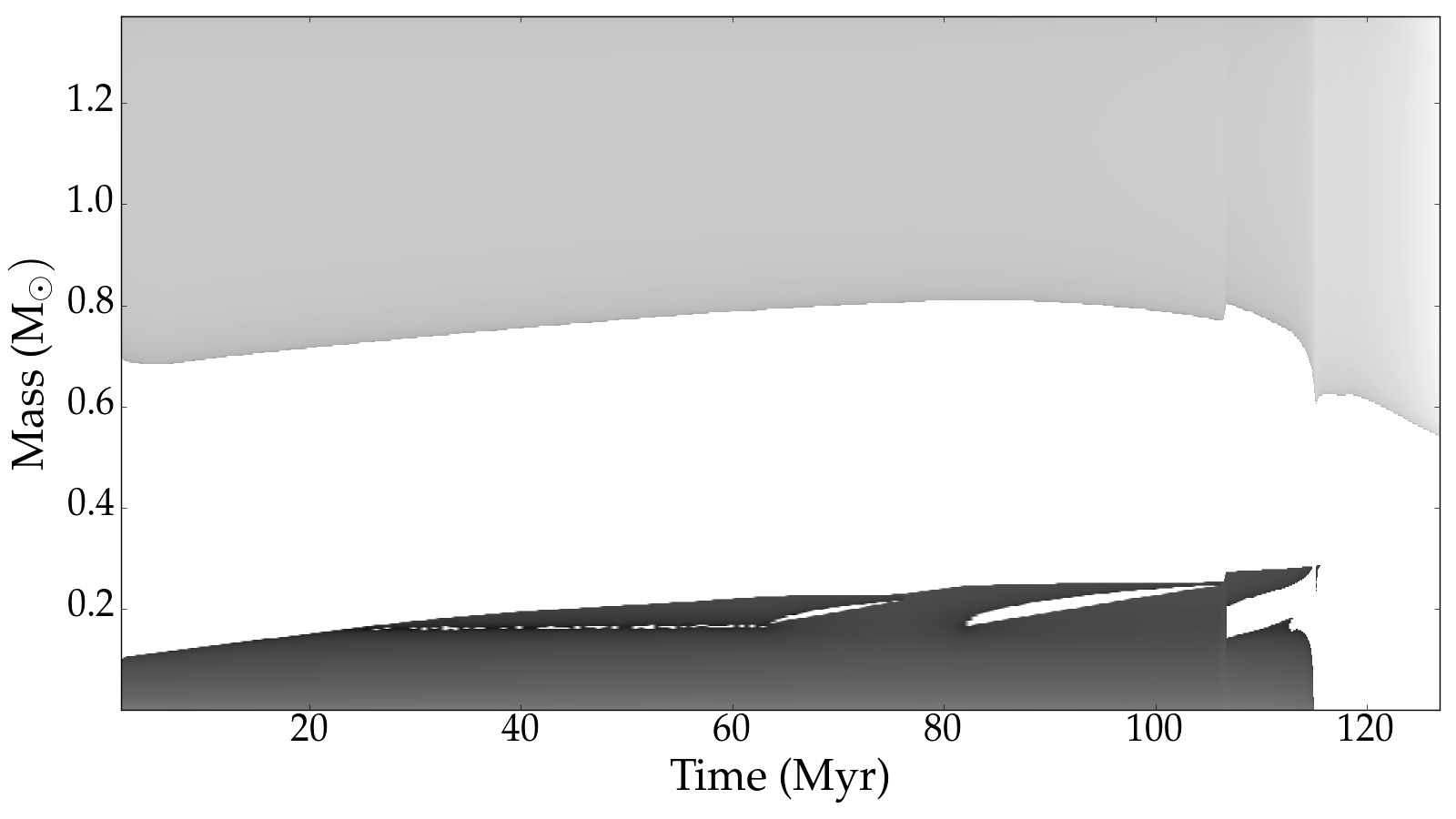}} & 
\subfloat[]{\includegraphics[width=80mm]{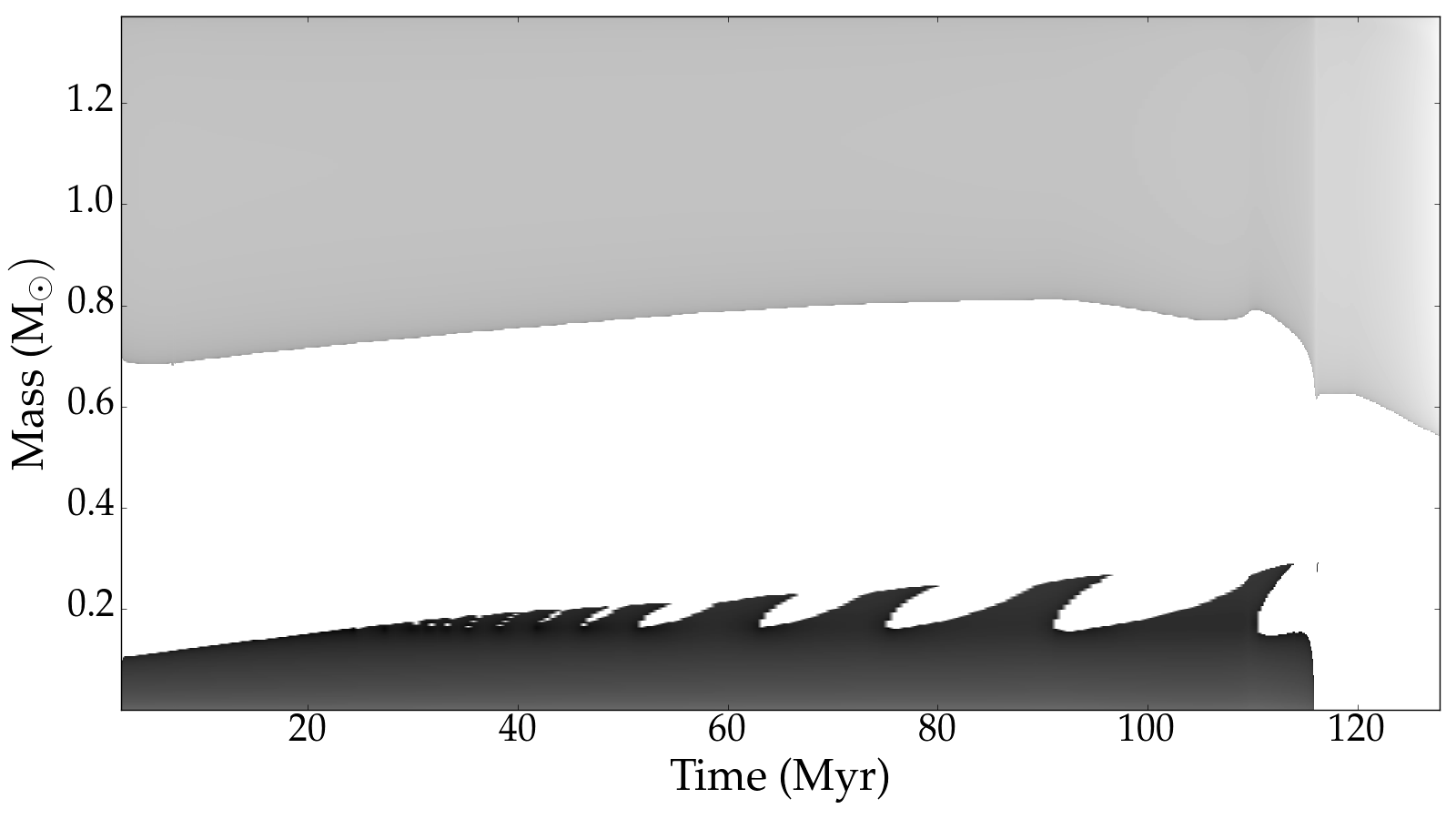}}\\
\subfloat[]{\includegraphics[width=80mm]{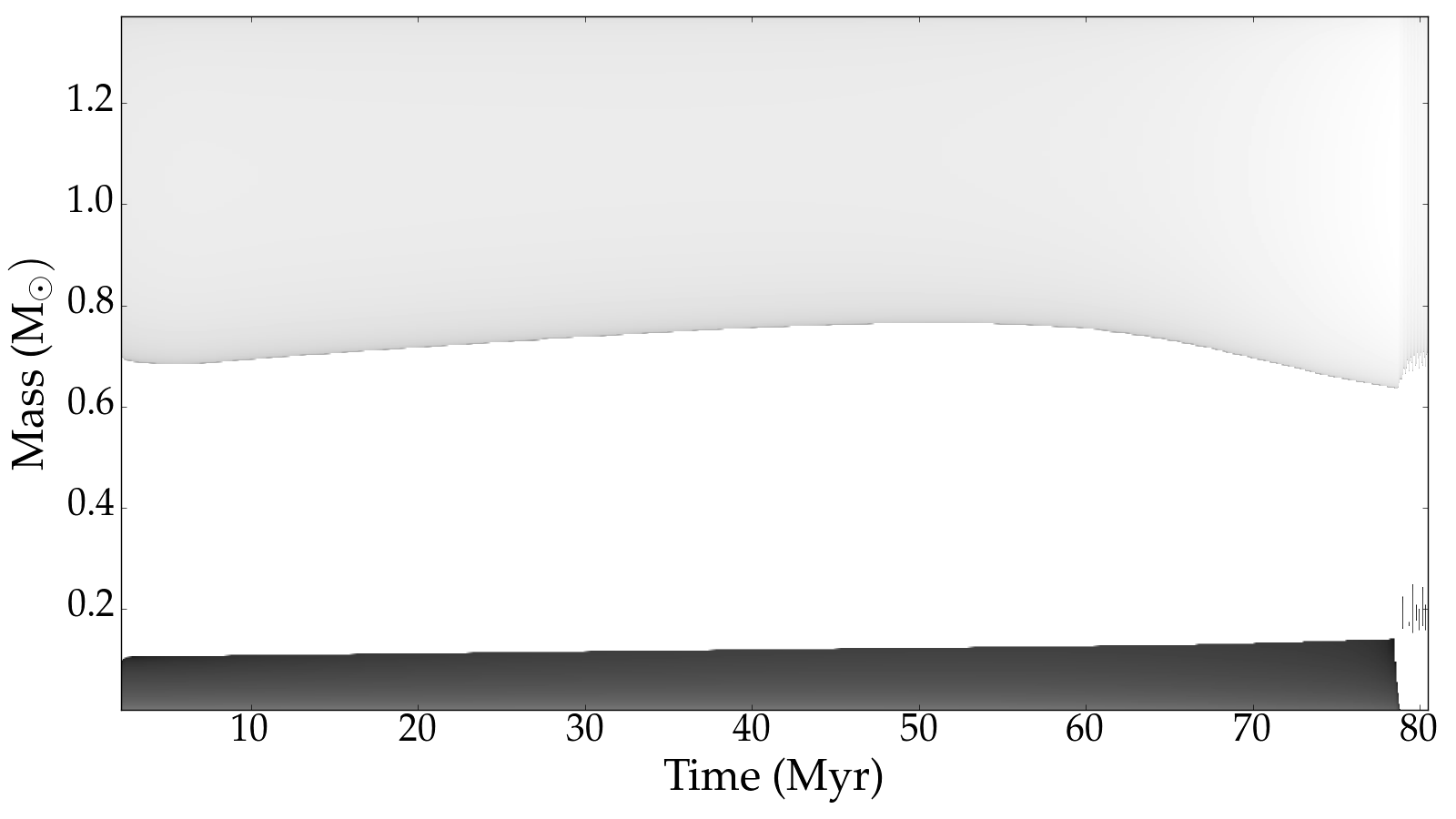}}&
\subfloat[]{\includegraphics[width=80mm]{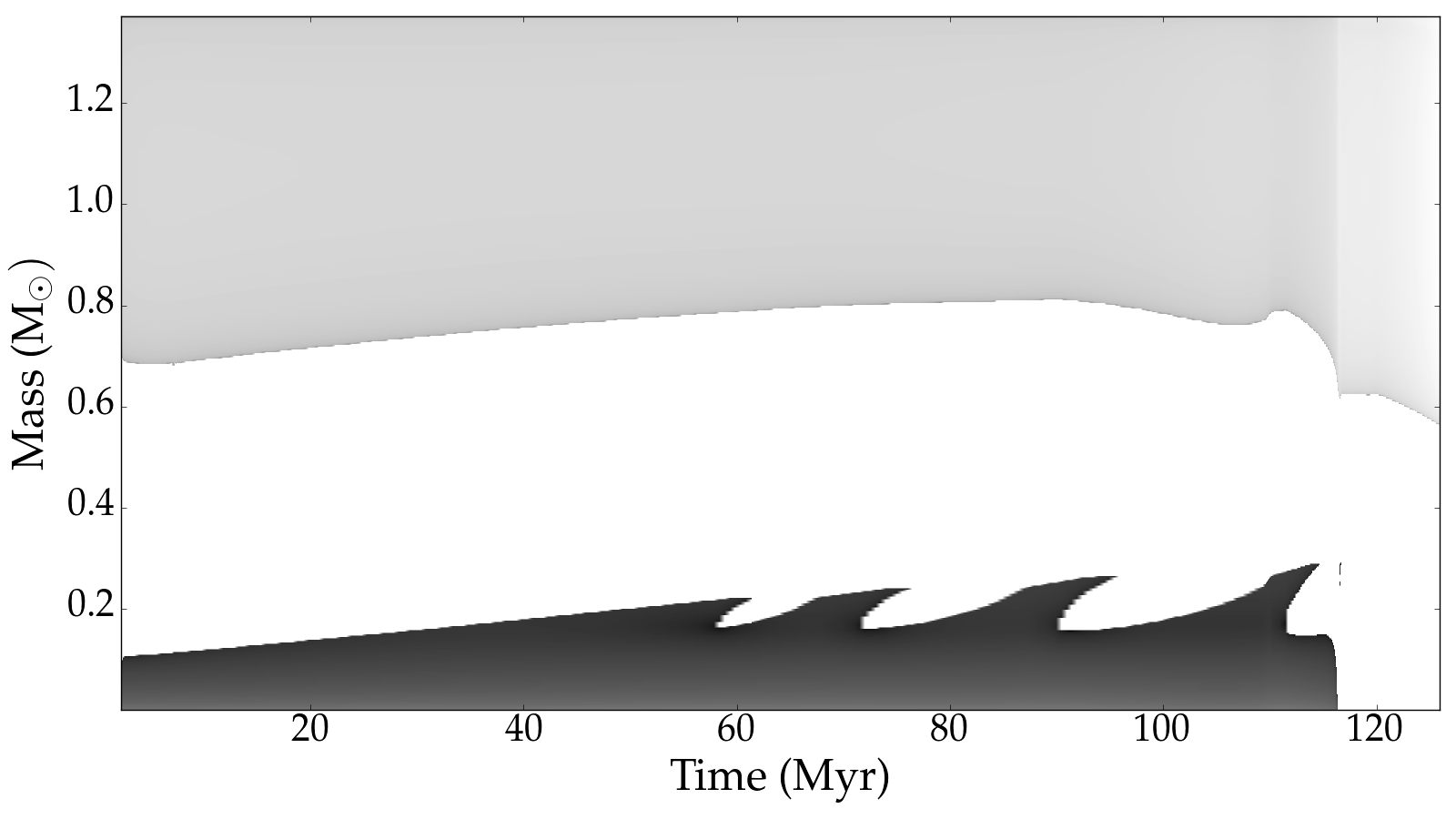}} \\
\end{tabular}
\caption{Convective core growing of HB stars as a function of star age for models with moderate (a), moderate plus diffusion (b), 
small (c), small plus diffusion (d), very small (e) and very small plus diffusion (f).}
\label{hb.png}
\end{figure*}

\begin{figure*}
\includegraphics[width=150mm]{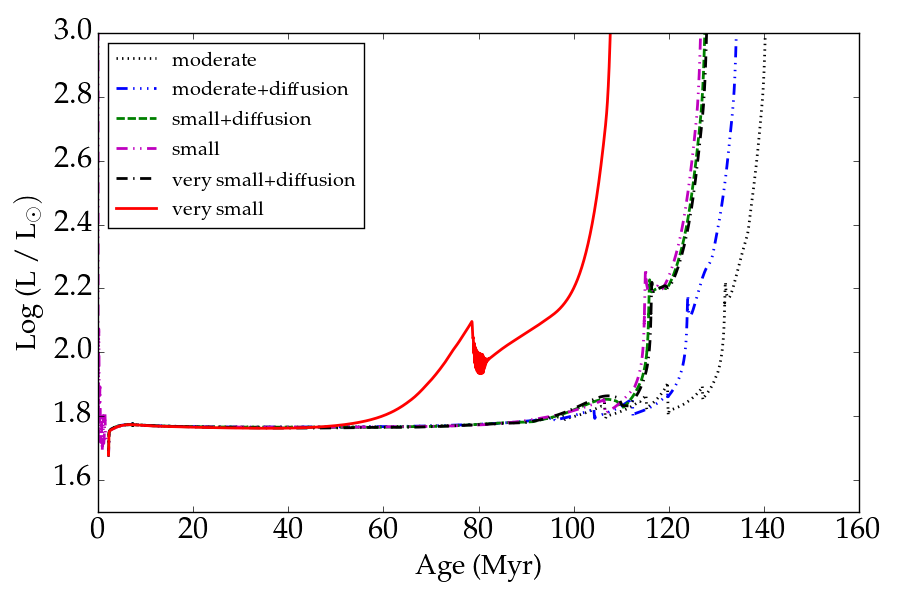}
\caption{Luminosity of a HB star as a function of  age, from the start of the helium flashes 
to the first thermal pulse on the asymptotic giant branch (TP-AGB).}
\label{r2.png}
\end{figure*}
\section{Discussion}

We investigated the influence of three overshooting scenarios on the evolution of
convective cores of sdB stars. In all mixing cases due to small and moderate overshooting, 
the accompanying mixing leads to fluctuations in the size of the convective core 
accompanied by the occurrence of convective shells during various stages of evolution.

The deviations from a constant g-mode period spacing are sensitive to the
detailed properties of the buoyancy frequency inside pulsating sdB stars. These
deviations can be interpreted by means of comparison with theoretical
calculations of the displacement eigenfrequencies and eigenfunctions. Our
results illustrate that a small amount of core overshooting,
accompanied by outer Brunt-V\"ais\"al\"a peaks due to convective shells are able
to produce the period spacing patterns observed in KIC\,10553698A at late stages. 
The Brunt-V\"ais\"al\"a peak of the transition layer between the He-rich envelope 
and the carbon core influences the trapping of modes and hence the period spacing patterns.

In MESA, a bare use of mixing length theory without additional mixing beyond the
convective core prohibits the core growth during EHB phase (see
\cite{2011ApJS..192....3P}, their Fig.\,15). Other stellar evolution codes rely
on an implementation of the mixing-length theory that allows for core growth
even for the cases without convective overshoot \citep{2014A&A...569A..63G,
  2015MNRAS.452..123C,2016MNRAS.456.3866C}.  Here, we have only considered MESA
models including convective overshoot as a phenomenon of convection carrying
material beyond the boundary of convective core; hence, the convective velocity
is non-zero at the core boundary and it declines exponentially with distance in
the overshooting layers.  \citet{2015MNRAS.452..123C} argued that the mode
trapping in KIC\,10553698A is a consequence of the sharp composition gradient at
the outer region of the convective boundary in their standard-overshoot mode
adopted, which has $f=0.001$. In their work, which is based on independent
evolution and pulsation codes, this amount of overshooting tends to grow the
core. At the outer edge of the convective core, the region becomes radiative.
The process continuously repeats, leaving behind a characteristic stepped
abundance profile.  It is not clear to us how their treatment of smoothing,
atomic diffusion, and the chosen time steps along the evolutionary track compare
to ours.  In their adopted moderate overshooting scenario, the convective core
growth explicitly prevented the splitting of the core and the authors were able to
produce the observed period spacing pattern of KIC\,10553698A.  In both our
moderate and small overshooting models, the appearance of convective shells
occurs and it also leads to the correct number of non-trapped modes between trapped
modes compared with the observations.  In addition, atomic diffusion smooths the
other Brunt-V\"ais\"al\"a peaks and leads to a better match with the
observations.
 
Horizontal Branch (HB) stars burn helium in their core and hydrogen in a
convective shell.  According to the Fig.\,~\ref{hb.png}, HB
convective core growing scenarios due to different amount of core overshooting are
almost similar to the EHB evolutionary scenarios. The $R_2$ parameter is the
fraction of asymptotic giant branch (AGB) stars to HB stars in globular
clusters \citep {1985A&A...145...97B,2001A&A...366..578C}.  Theoretical models allow 
to predict
$R_2$ from the duration of the various evolutionary phases and to consider
this as test that is very sensitive to the extension of the convective
core, which is indepedent of the seismic modelling of the period spacings. 
The $R_2$ parameter is less sensitive to other stellar quantities such as
initial mass and metalicity \citep {1989ApJ...340..241C}.  We computed the $R_2$
value for our models with convective core overshoot mixing due to moderate, small and very small
overshooting, with and without diffusion (see
Fig.\,~\ref{r2.png}).  Our results are listed in Table\,\ref{tab:table1}:  
for small overshooting and for very small overshooting in presence
diffusion,  $R_2=0.11$, which is compatible with recent observed values
\citep {2016MNRAS.456.3866C}.  Moreover, based on theoretical grounds,
\citet{1971Ap&SS..10..340C} proposed 
a very small overshooting layer because the high gravity in the core helium burning
phase acts as a braking force for convective motions inside the overshooting
layers. This is entirely compatible with our results.

\begin{table}
  \centering
  \caption{R2 parameters for models with different overshooting}
  \label{tab:table1}
  \begin{tabular}{cccc}
      & HeCB phase lifetime(Myr) & R2 parameter\\
    moderate &131.7  &0.07 \\
    small &114.8  &0.11 \\
    very small &79.1  &0.38 \\
    moderate+diffusion &123.8  &0.09 \\
    small+diffusion &115.7  &0.11 \\
    very small+diffusion & 116.2 &0.11 \\
%    \bottomrule
  \end{tabular}
\end{table}

An increased number of observed pulsating sdB stars with uninterrupted
high-precision space photometry will hopefully lead to an increased fraction of
sdB stars with trapped modes in the near future.  This would allow to
investigate the effects of time-dependent core overshooting on the size of the
convective core and g-mode behaviour of sdB pulsators. The overshooting distance
is inversely proportional to the differences of the mean atomic weight in the
interior and exterior of the convective core and depends on the gravitational
acceleration in the overshooting layers.  Matter exterior to the core is highly
buoyant relative to matter inside the core and only a fraction of additional
matter will be carried by the convective flows, as discussed in
\cite{1990ASPC...11....1S, 2015A&A...582L...2S}. Different cooling process of
overshooting layers, such as neutrino cooling, can be studied as a
time-dependent overshooting undoubtedly influencing the breathing pulses for the
late carbon burning phases \cite{2015ApJ...809...30A, 2015A&A...580A..61V}.
Generally, asteroseismic studies of g-mode pulsating stars, provide an excellent
opportunity to improve models of convection and overshooting in different stages
of stellar evolution, as indicated in the recent 321D algorithm development by
\cite{2015ApJ...809...30A}.

\section*{Acknowledgements}

The first author wishes to especially thank Roy {\O}stensen for interesting
discussions which helped to improve several parts of this study. The work was
initiated during a 9-months research stay at Leuven University. Part of the research stay of
HG at Leuven University was funded by the Francqui Prize offered to CA in 2012
by the Francqui Foundation, Belgium.  This research was also partly funded from
the European Community's Seventh Framework Programme FP7-SPACE-2011-1, project
number 312844 (SPACEINN).  The research leading to these results has received
funding from the People Programme (Marie Curie Actions) of the European Union's
Seventh Framework Programme FP7/2007--2013/ under REA grant agreement No. 623303
(ASAMBA).

\bibliographystyle{mnras}
\bibliography{Hamed}  % if your bibtex file is called example.bib

\bsp	% typesetting comment
\label{lastpage}
\end{document}